\documentclass[9pt,twocolumn]{extarticle}

\usepackage[letterpaper,margin=0.75in]{geometry}
\usepackage{times}        
\usepackage{amsmath,amssymb,amsfonts}
\usepackage{graphicx}
\usepackage{bm}
\usepackage[normalem]{ulem}   
\usepackage{xcolor}
\usepackage{authblk}      
\usepackage{titlesec}     
\usepackage{abstract}     
\usepackage{caption}      

\usepackage{float}
\usepackage[most]{tcolorbox}

\titleformat{\section}
  {\large\bfseries\uppercase}{\thesection.}{1em}{}
\titleformat{\subsection}
  {\normalsize\bfseries}{\thesubsection.}{0.5em}{}
\titleformat{\subsubsection}
  {\normalsize\itshape}{\thesubsubsection.}{0.5em}{}

\titlespacing*{\section}{0pt}{1.1em}{0.5em}
\titlespacing*{\subsection}{0pt}{1em}{0.25em}

\newcommand{\keywords}[1]{%
  \vspace{3pt}
  {\noindent\small{\bf Keywords:} #1}
}

\title{\textbf{Cell Deformation Signatures along the Apical-Basal Axis: \\
       A 3D Continuum Mechanics Shell Model}}

\author[1,2]{\small Jairo M. Rojas}
\author[1,2]{\small Mayisha Z. Nakib}
\author[3]{\small Vivian W. Tang}
\author[3]{\small William M. Brieher}
\author[1,2]{\small Sascha Hilgenfeldt}
\affil[1]{\footnotesize Department of Physics, The Grainger College of Engineering,
  University of Illinois, Urbana-Champaign, IL 61801, USA}
\affil[2]{\footnotesize Mechanical Science and Engineering,
  The Grainger College of Engineering, University of Illinois,
  Urbana-Champaign, IL 61801, USA}
\affil[3]{\footnotesize Department of Cell and Developmental Biology,
  University of Illinois, Urbana-Champaign, IL 61801, USA}

\date{}  

\begin{document}
\twocolumn[   
    \maketitle

  \vspace{-2em} 
    \begin{onecolabstract}
    \noindent
Two-dimensional (2D) mechanical models of confluent tissues have related the mechanical state of a monolayer of cells to the average perimeter length of the cell cross sections, predicting floppiness or rigidity of the material. For the well-studied system of in-vitro MDCK epithelial cells, however, we find experimentally that cells in mechanically rigid tissues display long perimeters characteristic of a floppy state in 2D models. We suggest that this discrepancy is due to mechanical effects in the third (apical-basal) dimension, including those caused by actin stress fibers near the basal membrane. To quantitatively understand cell deformations in 3D, we develop a continuum mechanics model of epithelial cells as  elastic cylindrical shells, with appropriate boundary conditions reflecting both the passive confinement of neighboring cells and the active stress of actomyosin contractility. This formalism yields analytical solutions predicting cell cross sections along the entire cylinder axis. Deconvolution microscopy experimental data confirm the significant and systematic change in cell shape parameters in this apical-basal direction. In addition to providing a wealth of detailed information on deformation on the subcellular scale, the results of the approach alter our understanding of how active tissues balance requirements of their stiffness and integrity, suggesting they are more robust against loss of rigidity than previously inferred.
    \end{onecolabstract}
    \vspace{0.5cm}
]

\begin{tcolorbox}[title=Significance Statement, colback=gray!10, colframe=black]
Linking the morphology of cellular tissue to its mechanical properties is invaluable for understanding functionality and diagnosing pathological changes  simply by imaging the tissue. Existing models connect two-dimensional statistical patterns with the mechanical energy of a cell monolayer. The present work shows that such 2D modeling is not quantitative for common epithelial tissues because actin cell mechanics varies characteristically along the apical-basal axis perpendicular to the monolayer. Instead, an entirely new approach to modeling is accomplished, describing cells as three-dimensional elastic shells  and arriving at a quantitative description of the experimentally observed morphologies. This theory of tissue mechanics provides richer detail for comparison with biological experiments on the sub-cellular level as well as a new paradigm for larger-scale tissue diagnostics.
\end{tcolorbox}

\section*{}

All metazoan organisms build interconnected sheets of epithelial cells to separate two different environments. While the sheet structure has to change flexibly during development, the final epithelium must maintain its integrity robustly throughout life to act as a reliable barrier despite being made from thousands to millions of individual cells. Much work has been devoted to elucidate tissue development, but our understanding of the robust equilibrium state of mature tissue is rudimentary. We know in some detail which biomolecules are necessary for building and maintaining epithelial sheets, but we do not understand how the molecular foundations integrate in space and time to the multicellular length scale and life time of the stable tissue. In particular, a fundamental understanding of mechanical properties of tissue as a cellular material has lagged behind our knowledge of genetics and regulatory pathways. This is because tissues are soft, nonlinear, complex materials made from typically disordered arrangements of cells. Only recently, mathematical tools have been developed that make use of the cellular structure in order to relate geometry, topology, and cell size statistics to the large-scale mechanical behavior of the tissue, and by extension to its biological function. The ultimate goal is to diagnose the health and functionality of a tissue through simple snapshots of its structure, extracting morphological indicators that quantitatively relate to mechanics.

The simplest type of such a modeling approach describes a monolayer of cells (often an epithelium) as a two-dimensional (2D) material made of confluent polygonal domains that share one-dimensional edges meeting in vertices. The vertex positions are used as degrees of freedom in a large class of models (vertex models), where mechanical energy can be associated with perimeter elasticity, area elasticity, and cell-cell adhesion \cite{Staple10,Bi_2015,Kim15}. A finite continuous phase between cells can also be taken into account \cite{kim2021embryonic,banavar2021mechanical}. In a pioneering success of vertex models, a transition from rigid ground states (finite shear modulus of the tissue) to floppy ground states (vanishing shear modulus) was demonstrated \cite{Staple10,Bi_2015}, a phenomenon for which area elasticity is not important and which, for fixed areas, is nearly independent of polydispersity of areas \cite{Kim15}. A main morphological indicator for tissue stiffness was found to be the perimeter length of the cell outlines normalized by the square root of cross-sectional areas, either as shape index \cite{Bi_2015} or as equivalent foam energy \cite{Kim19}.

Using perimeter lengths as quantitative indicators of mechanics requires care, however, and a number of modifications have been suggested to pinpoint the boundary between rigid and floppy states, changing quantitative predictions depending on cell topology or overall anisotropic stress \cite{wang2020anisotropy,thomas2023shape}. Moreover, the original idea of a single critical perimeter value indicating a transition of the mechanical ground state has been augmented to account for metastable states on an energy landscape \cite{Kim19,Kim22}, where mechanically rigid states can display a range of perimeter measure values. We shall demonstrate, however, that common, perfectly rigid epithelial tissue systems do not conform to even the broadest interpretation of these theoretical predictions when their morphology is analyzed and quantified.

\begin{figure}
    \includegraphics[width=0.45\textwidth]{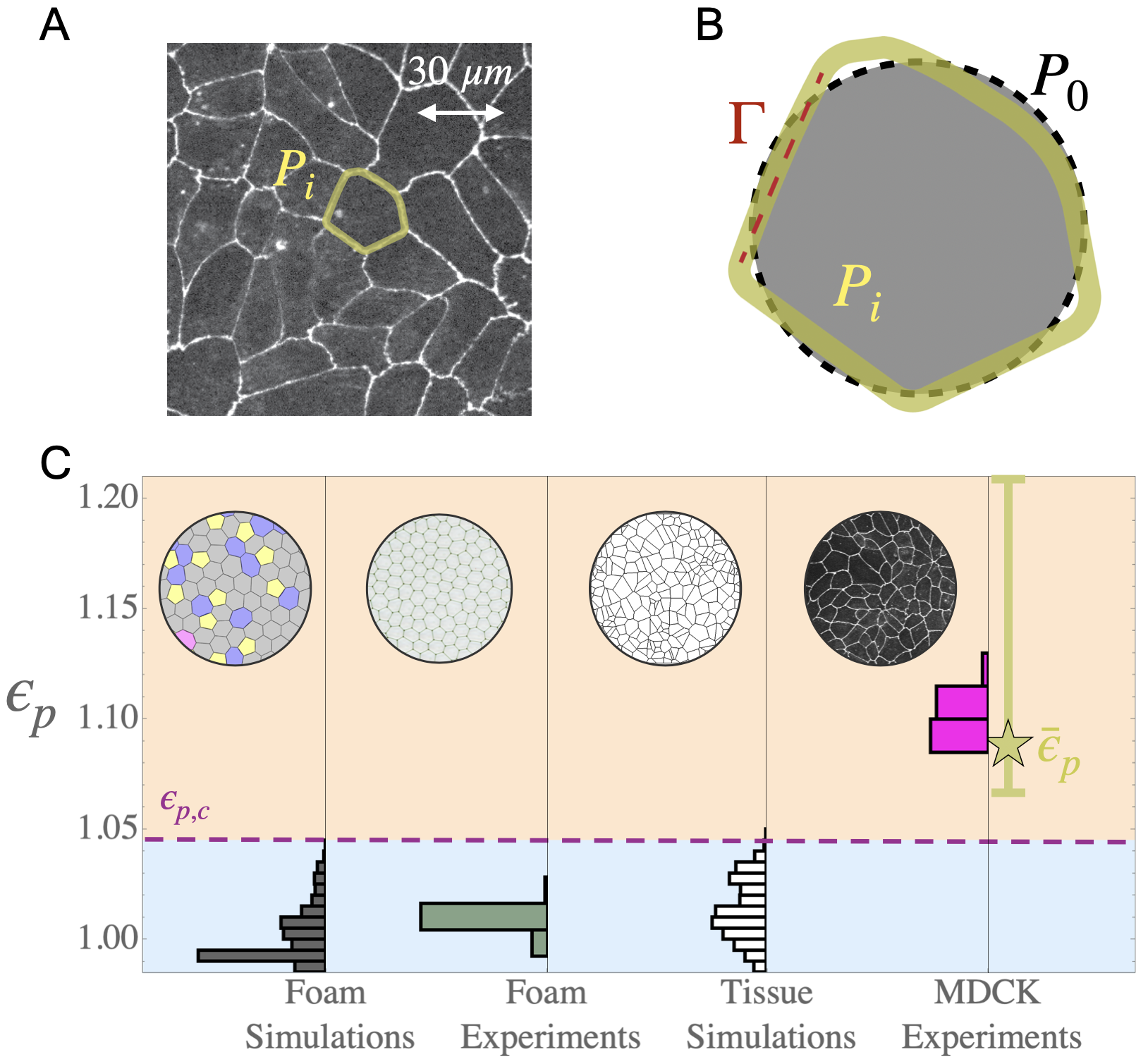}
    \caption{(A) MDCKII tissue sample (fluorescent signal: tight junction marker ZO-1 \cite{anderson1988characterization}), with perimeter length of a cell indicated. (B)  Two-dimensional tissue morphology: cell perimeter $P_i$ deviates from equilibrium perimeter $P_0$, giving rise to an elastic energy $\propto (P_i-P_0)^2$. Along $P_i$ in the confluent tissue, a specific adhesion energy $\Gamma$ is active. (C) Two-dimensional perimeter measures according to \eqref{epsilonp} for experimental and simulated samples of foams and tissues. The purple dashed line marks the largest value $\epsilon_{p,c}$ compatible with rigid behavior according to 2D vertex theories. Yellow line and star are the range and mean of the $z$-resolved data in Fig.~\ref{fig:Perimeter Measurement.png}.}
    \label{fig:2Dresults}
\end{figure}

Madin-Darby canine kidney (MDCK) cells were maintained at confluence for 10 days on transwell filters to ensure that the cells were polarized and that adhesive junctions were mature.  Cell boundaries were identified by immunofluorescence in fixed cells using deconvolution microscopy to collect Z-stacks of images from the apical tight junction to the basal surface, which was identified by the appearance of actin stress fibers. Occludin or ZO-1 were used to mark tight junctions, E-cadherin or beta-catenin to mark the lateral membranes, and phalloidin to stain actin filaments. Cell outlines were extracted by automatic segmentation (see Materials and Methods for details). 
Figure~\ref{fig:2Dresults}A shows such a sample and an extracted perimeter $P_i$
of a single cell of cross section $A_i$. Entire samples of $N\sim 2000$ confluent cells were analyzed in order to quantify a normalized perimeter measure,
\begin{equation}
\epsilon_p = \frac{1}{N P_{hex}} \sum_{i=1}^N P_i\,,
    \label{epsilonp}
\end{equation}
using the perimeter $P_{hex}$ of a regular hexagon of area equal to the average cross-sectional area $A_0\equiv \sum A_i / N$. This measure is equivalent to the shape index \cite{Bi_2015}, but is more easily interpreted: The $\epsilon_p$ of confluent polygonal patterns is bounded from below by 1 if monodisperse (the regular honeycomb), or by a value slightly below 1 if moderately polydisperse \cite{Kim19}. Note also that even regular hexagons are elastically strained in common 2D vertex theories, as the unstrained reference perimeter $P_0$ is smaller than $P_{hex}$, here taken to be circular (Fig.~\ref{fig:2Dresults}B). The elastic energy is lowered by cell-cell adhesion of specific energy $\Gamma$ along the perimeter. Figure~\ref{fig:2Dresults}C shows distributions of $\epsilon_p$ for 2D data from many samples, not just for MDCK cells, but for experiments with single layers of foam bubbles as well as for simulations of 2D tissue and foam systems. All of these systems are rigid, i.e., possess a finite shear modulus tested in either experiment or simulation. As the simplest and most robust prediction for such systems, the figure indicates the maximum value of $\epsilon_{p,c}$ compatible with rigid behavior (configurations with $\epsilon_p>\epsilon_{p,c}$ are predicted to be "floppy", i.e., can be strained without expense of mechanical energy). The simulations and the foam experiments conform to this prediction, while the experimental MDCK data deviate strikingly: none of the samples is in the predicted rigid range, displaying much larger perimeter measures, despite being manifestly rigid in direct mechanical tests, which show that the epithelial sheet has a finite elastic modulus and tears if too much force is applied \cite{schulze2017elastic,tang2013fsgs3}.

\begin{figure}
    \centering
    \includegraphics[width=0.45\textwidth]{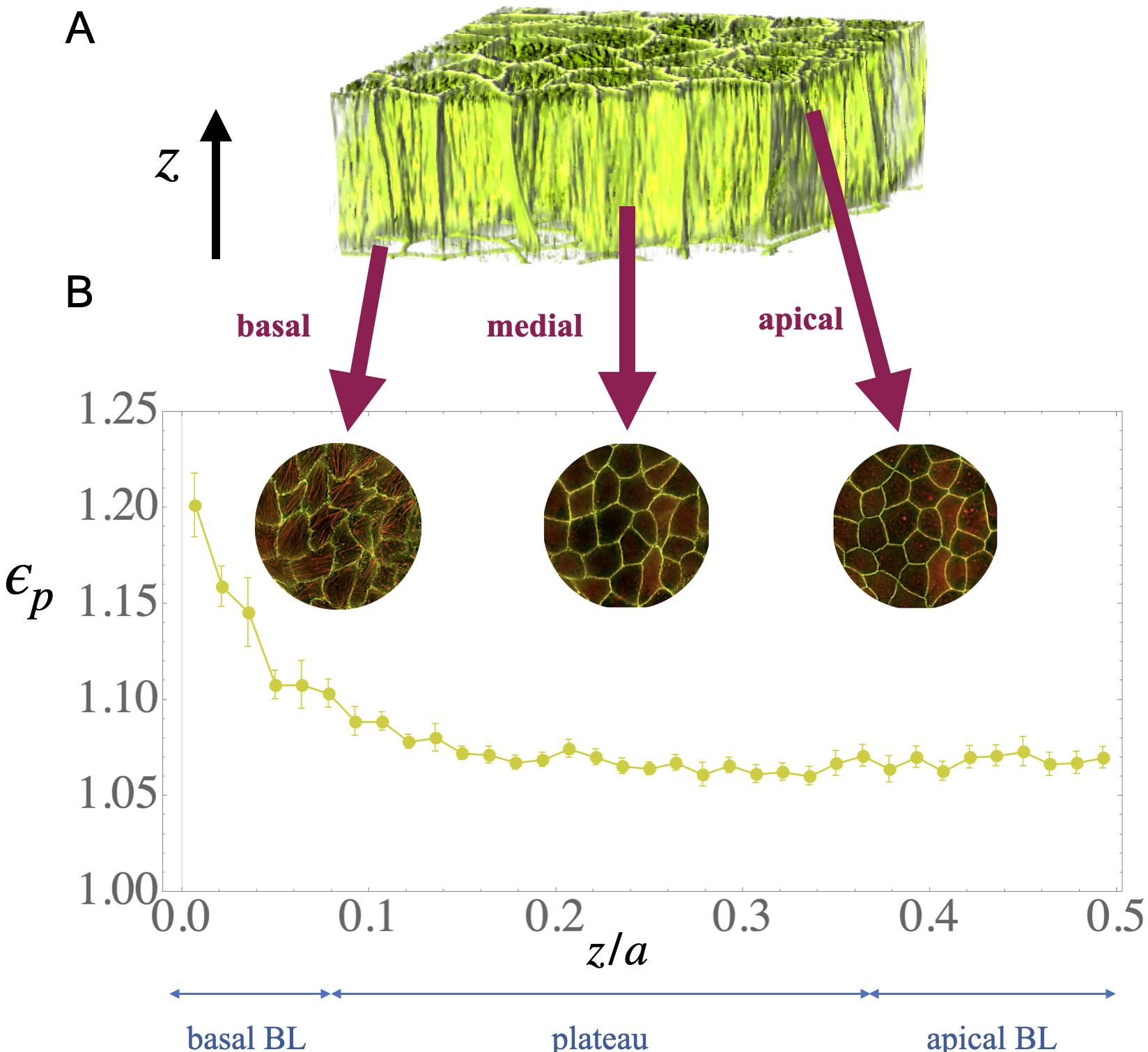}
    \caption{Perimeter measures resolved along the $z$-axis. In several stacks of deconvolution microscopy images spanning the entire epithelial height (An example of a 3D reconstruction is shown in A), \eqref{epsilonp} is computed for every slice and plotted against $z/a$ with the mean cell radius $a=15\mu$m (B). All values are significantly above  $\epsilon_{p,c}$, and a prominent boundary layer is observed at the basal ($z=0$) end.
    }
    \label{fig:Perimeter Measurement.png}
\end{figure}

\subsection*{Optical sectioning images reveal 3D shape variations}

Images of smaller samples of the same type of MDCKII tissue were taken with higher magnification (100x) using widefield imaging and deconvolution microscopy allowing for submicron resolution along the apical-basal axis (the $z$-axis perpendicular to the 2D plane). These images reveal significant systematic changes in cross-sectional morphology as $z$ varies: while the neighbor topology of each polygonal cross section typically remains unchanged (scutoids are rare), the shape varies from nearly straight-edged polygons apically to polygons with edges that appear strongly curved or buckled (cf.\ insets in Fig.~\ref{fig:Perimeter Measurement.png}). In parallel, the perimeter measure $\epsilon_p$ changes systematically with $z$, showing a prominent boundary layer with high and strongly varying values at the basal end and a 
nearly constant plateau value over $\sim 70\%$ of the cell height. Figure~\ref{fig:Perimeter Measurement.png}B quantifies this trend using averaged perimeter data from six 100x magnification samples of 15-20 cells each. We normalize the $z$-axis by the effective radius $a\approx 15\mu$m following from the average cross-sectional area $A_0=\pi a^2$. Individual samples, and even the shapes of individual cells, robustly show the same behavior as Fig.~\ref{fig:Perimeter Measurement.png}B, see Supporting Information (SI). A 2D model assuming that every cross section perpendicular to the apical-basal axis shows the same structure is clearly not adequate, rationalizing the failure to model lower-resolution data in 2D. It is important to note that even the plateau value of Fig.~\ref{fig:Perimeter Measurement.png}B is significantly higher than $\epsilon_{p,c}$, so that the effect of 3D variation is not local. The height-averaged perimeter $\bar{\epsilon}_p\approx 1.09$ is consistent with the  experimental measurements of Fig.~\ref{fig:2Dresults}C, whose low resolution results in an effectively averaged signal over $z$.

The 100x images also reveal a markedly heterogeneous distribution of actin in general and actin fibers in particular. As actomyosin contractility is thought to be the main driver of active cell and tissue mechanics \cite{lecuit2011force,heisenberg2013forces}, this provides direct clues as to the distribution of mechanical stresses along the $z$-axis. At the basal side of the tissue ($z\approx 0$), prominent actin bundles \cite{tang2012alpha,saraswathibhatla2020tractions,jetta2023epithelial} traverse each cell with a distinct orientation for each cell, resulting in strongly anisotropic stresses. The coupling of the forces from these bundles to the cellular cortex gives rise to deformations not just at $z=0$, but throughout the height of the cells. Similarly, a more diffuse and isotropic  distribution of branched actin \cite{tang2013fsgs3,tang2012alpha} is seen at the apical side of each cell ($z\approx H$). In agreement with literature \cite{martin2009pulsed,smutny2010myosin}, we see less actomyosin activity along the lateral sides of each cell; we will thus conceptualize active  stresses as being applied at the basal and apical ends of a cell. 

\begin{figure*}
    \centering
    \includegraphics[width=0.80\textwidth]{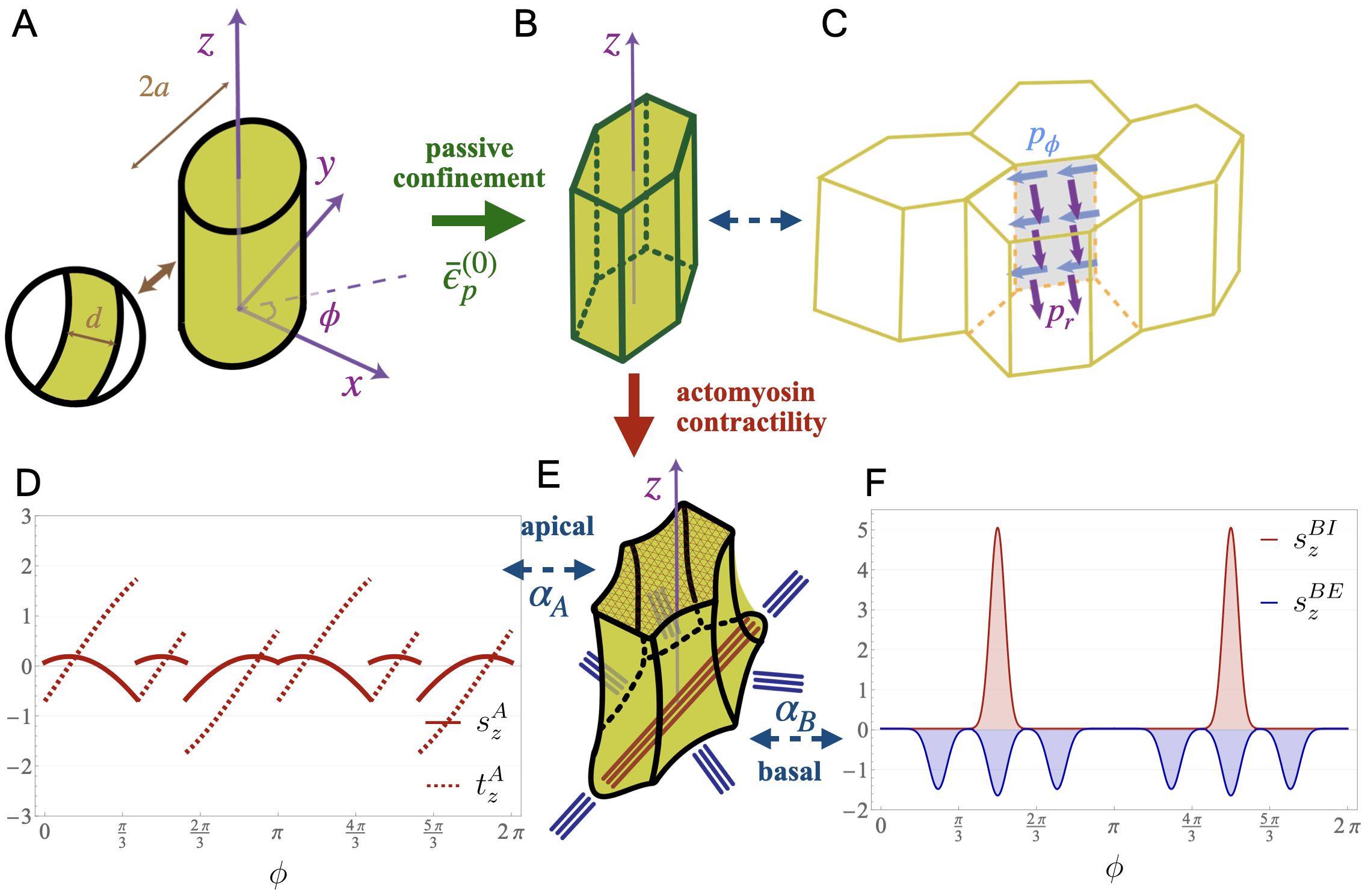}
    \caption{Elastic-shell model for 3D deformations of epithelial cells. (A) Circular cylinder shell as undeformed reference. (B) Deformation through passive confinement by the stresses $p_r, p_\phi$ from neighboring cells (C) results in a polygonal cylinder with perimeter measure $\bar{\epsilon}_p^{(0)}$. Active stresses are applied according to the observed actomyosin distributions on the apical (D) and basal (F) sides with strengths $\alpha_A$ and $\alpha_B$, respectively, resulting in the final deformed morphology (E).
    }
    \label{fig:Cell Diagram 2}
\end{figure*}

\section*{3D Shell Model}

In order to make use of the rich information of apical-basal variation of activity and in turn quantitatively describe the entire cell morphology, we need to formulate an augmented model in three dimensions. Existing approaches almost exclusively adopt a 3D vertex model with polyhedral cells \cite{alt2017vertex,rozman2020collective}. However, such models have many free parameters and must make ad-hoc assumptions about the mechanical response of the faces and edges of the polyhedron depending on their orientation. They also inherently do not have the resolution to describe continuous changes along the apical-basal axis, which figure prominently in our data (Fig.~\ref{fig:Perimeter Measurement.png}). Therefore, we here introduce a radically new concept and treat the entire height of the cell as an elastic continuum shell deformed by (i) the passive confinement of the neighboring cells and (ii) the active stresses applied by actin basally and apically (Fig.~\ref{fig:Cell Diagram 2}). This severely cuts down on free parameters and in fact relates all parameters directly to experimental observations or mechanical tests. The elastic-shell model provides an in-detail description of global 3D morphology of a cell, and consequently opens up a rich space of more sophisticated morphological indicators for inference of tissue mechanics.

\subsection*{Elastic Shell Theory}
Restricting ourselves to thin shells of linear elastic material (Young's modulus $Y$ and Poisson ratio $\nu$) is a severe simplification that disregards dissipation, viscoelasticity, and nonlinearity, but it leads to a complete description of the theory's solution space and is conceptually appropriate for modeling an equilibrium, robust state of a rigid tissue to leading order. As the prediction of shape is the main outcome of our approach, we adopt continuum equations for the displacements of shell points in a cylindrical $(r,\phi,z)$ coordinate system (Fig.~\ref{fig:Cell Diagram 2}A). Shell theory averages across the thickness $d$ of the shell, leaving only $z$ and $\phi$ as independent variables. The unstrained shape of the cell is an upright circular cylinder mantle of radius $a$ and height $\Lambda a$. We shall normalize all lengths by $a$, resulting in dimensionless shell thickness $t=d/a$ and non-dimensional displacement measures $(u,v,w)$ in the $(r,\phi,z)$ directions, respectively. The linearity of the system of equations (see Methods) allows for a decomposition into Fourier modes of the $\phi$ dependence. Our goal is to model a mean cell shape, thus retaining mirror symmetry of cross-sectional cell shape with respect to the $x$- and $y$-axes. Therefore, only even modes with dependences $\cos(2n\phi)$ (or $\sin(2n\phi)$ for derived quantities) need to be considered. For example, the radial displacement function encoding for perimeter shape changes is decomposed as $w(z,\phi) = \sum_n w_n(z) \cos(2n\phi)$.

\subsection*{Boundary Conditions of passive confinement} For each mode $n$, the system of equations can then be rewritten as ODEs for $u_n(z)$, $v_n(z)$, and $w_n(z)$.

Each $w_n(z)$ has a particular solution and eight linearly independent complex-exponential homogeneous solutions (see Methods for details). The particular solutions are determined by the Fourier components of external normal and tangential forces $p_r$ and $p_\phi$ acting on the mantle area (see Fig.~\ref{fig:Cell Diagram 2}C). These forces represent the passive confinement by the neighboring cells in the confluent tissue and are chosen such that the particular solution is a cylinder with polygonal cross section (Fig.~\ref{fig:Cell Diagram 2}B) whose perimeter $\bar{\epsilon}_p^{(0)}$ is consistent with the rigid range of 2D vertex theory shaded blue in Fig.~\ref{fig:2Dresults}C, i.e., $1 \leq \bar{\epsilon}_p^{(0)} \leq \epsilon_{p,c}$. 

\begin{figure*}[h]
    \centering
    \includegraphics[width=1.0\textwidth]{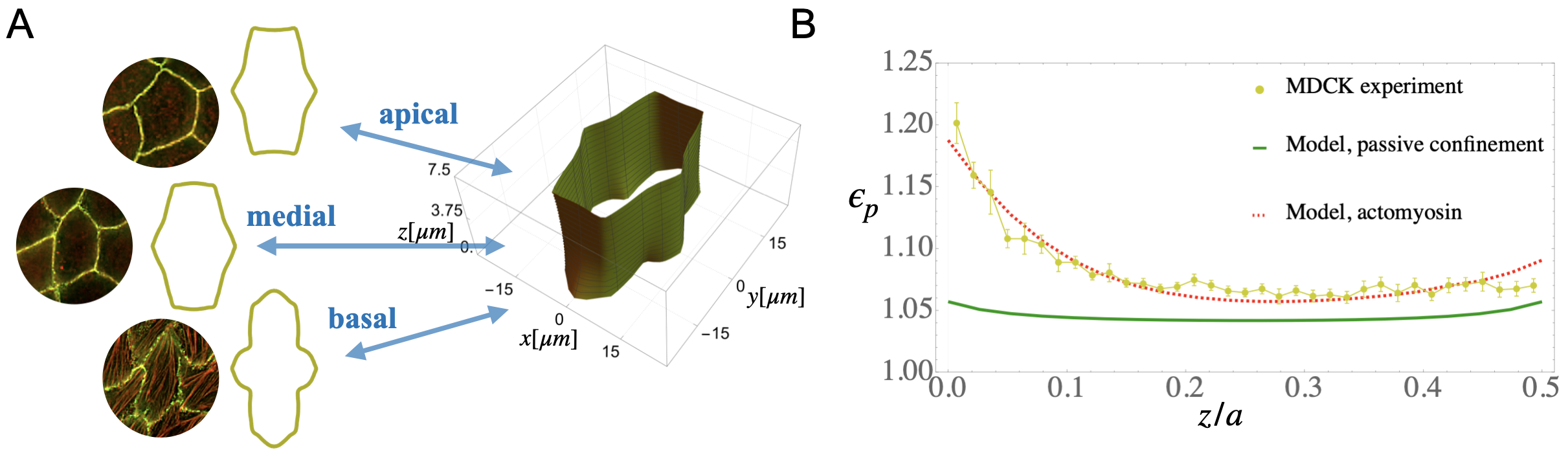}
    \caption{Modeling results. (A) 3D shape deformation of the cylindrical shell, with cross sections at $=0,\Lambda a/2,\Lambda a$ indicated on the left and compared with experimental cross sections of the same cell in image stack layers at the same coordinates. (B) Model-derived apical-basal dependence of normalized perimeter $\epsilon_p(z)$ is in very good agreement with experiments. Modeling with only passive deformation (no actomyosin contractility, green) fails to raise the perimeter beyond the 2D theory range (cf.~Fig.~\ref{fig:2Dresults}. The shell results presented here are based on the largest passive deformation compatible with rigid 2D shapes, i.e.,  $\bar{\epsilon}_p^{(0)}=\bar{\epsilon}_{p,c}$.
    }
    \label{fig:Cell Results}
\end{figure*}

\subsection*{Boundary Conditions of Actin stress}
The homogeneous solutions yielding the full three-dimensional, $z$-dependent deformation of the cylinder (e.g.\ quantified by $\epsilon_p(z)$) are informed by the boundary conditions at the basal ($z=0$) and apical ($z=\Lambda a$) ends, see Methods. Basally, no-displacement conditions $u=v=0$ ensure connection to the substrate. Additionally, resultants of basal active stress are inferred 
directly from the experiments:  we observe a strongly directional actin bundle in each cell, with nearly random directions uncorrelated from cell to cell (see SI). These bundles exert normal stress on the basal cortex, translated to effective radial shear stress resultants $S_z$ \cite{flügge2013stresses} at the lower edge of the cylinder. The azimuthal distribution of $S_z^B(\phi)=S_z(z=0,\phi)$ is taken to reflect the pattern of such bundles, adding the stress $S_z^{BI}$ of the interior bundle of the modeled average cell (taken to point in the $\phi=\pi/2$ direction by symmetry) to that of the exterior bundles of neighbors $S_z^{BE}$ (on average, every polygonal edge is impacted by 1/3 of an equivalent bundle force). The resulting boundary condition  $s_z^B(\phi)=s_z^{BI}+s_z^{BE}$ is shown in Fig.~\ref{fig:Cell Diagram 2}F, with each bundle a Gaussian peak in $\phi$ 
and using lowercase $s_z$ to indicate normalization to unit integral strength (see SI for details).  To complete the boundary conditions at $z=0$, we impose zero moment around the $z$-axis ($M_z=0$), as the average cell will not bend in any direction. 

The actin distribution at the apical side, by contrast, is a nearly uniform and isotropic cortex \cite{klingner2014isotropic}, see Fig.~\ref{fig:Perimeter Measurement.png}. The uniform boundary-normal stress causes both radial and tangential stress resultants at $z=\Lambda a$, denoted by $s_z^A(\phi)$ and $t_z^A(\phi)$, respectively, when normalized to unit total integral strength. The resulting patterns are illustrated in Fig.~\ref{fig:Cell Diagram 2}D, see SI for details. As the apical side of the cell is otherwise free, the remaining boundary conditions are homogeneous in bending moments and axial stress. 
Finally, the dimensionless (normalized) functions $s_z^B$ and $s_z^A,t_z^A$ are multiplied by actin strength factors $\alpha_B$ and $\alpha_A$, respectively. The final shape of the shell (Fig.~\ref{fig:Cell Diagram 2}E) is fully determined by the parameters $\bar{\epsilon}_p^{(0)}$, $\alpha_A$, and $\alpha_B$.

In this fashion, we have translated observed actin distributions and accompanying actomyosin contractility into a well-defined shell mechanics problem that can be solved analytically mode by mode by Fourier-decomposing the boundary conditions. Approximate values for the shell parameters are taken from the literature: An effective thickness of $H \approx 430$nm is common for epithelial cells given the $\approx 19$\,nm long cadherin ectodomain \cite{boggon2002c}, 19\,nm to bridge the cadherin-catenin complex to actin cytoskeleton \cite{ishiyama2012three}, and the cortical actin thickness, which is estimated at 190\,nm in rounded mitotic cells but closer to 400\,nm in isolated cells in interphase (see \cite{chugh2017actin}). Altogether, this translates to $t\approx 0.04$. The Young's modulus of an epithelial cortex is approximately $Y\approx 33$\,kPa \cite{schulze2017elastic},
and we assume a generic (3D) Poisson's ratio of $\nu=1/3$, completing the definition of modeling parameters.

\subsection*{Solutions}
In most engineering applications, the thin-shell equations \eqref{shellequations1}--\eqref{shellequations3} (Methods) can be approximated by lower-order systems, mostly relying on limits of very elongated cylinders (allowing for no exponentially growing solutions) or on squat cylinders (favoring nearly uniform solutions), cf.\ \cite{flügge2013stresses,mahadevan2007persistence}. 
While these may be appropriate simpifications for columnar and squamous epithelia, respectively, the  MDCK II cells in our case are approximately cuboidal (average aspect $\Lambda\approx 0.5$), and the full solutions apply. Thus, every deformation mode combines boundary-layer displacements and nearly uniform solution components (see SI for details). When composed from the weighted modes, the complete solution retains this character. 

In order to model the experimental shapes quantified in Fig.~\ref{fig:Perimeter Measurement.png}, the passive deformation by the neighbors $\bar{\epsilon}_p^{(0)}$ must be specified. In order to be as close to putative results of the 2D vertex theory as possible, we first choose the largest perimeter compatible with it, i.e., $\bar{\epsilon}_p^{(0)} = \epsilon_{p,c}$. The green curve of Fig.~\ref{fig:Cell Results}B shows the resulting (nearly uniform) perimeter profile with passive deformation only. To incorporate the active stresses, we perform a least-squares fit to the $\epsilon_p(z)$ shape of Fig.~\ref{fig:Perimeter Measurement.png}B, resulting in $\alpha_A^{(1)}\approx 2.3\times 10^{-4}, \alpha_B^{(1)}\approx 5.6\times 10^{-4}$. Note that these actin strengths are of comparable magnitude, but their very different morphology (diffuse branched network vs.\ fiber bundles, Fig.~\ref{fig:Cell Diagram 2}D,F) lead to strongly differing deformations near the basal and apical ends. 
The basal stress obtained through these model results can be rationalized through the resulting deformation at the basal end of the shell: assuming the basal cortex of the shell to be bonded firmly to the basal fibers attached to focal adhesions, the maximum radial shell strain $\varepsilon_r \sim 0.45$ at $\phi=\pi/2$ (location of the internal fiber bundle) is approximately equal to 2/3 of the internal-bundle fiber strain. Using the typical fiber Young's modulus of $Y_F \sim 300$\,kPa \cite{deguchi2005evaluation,kassianidou2015biomechanical} and the typical $r \sim 100$\,nm radius of fibers in epithelial cells  \cite{katoh1998isolation,kassianidou2015biomechanical}, we find the observed strain compatible with $ Y_F \pi r^2 (3/2)\varepsilon_r\sim 6.4$nN force per fiber, a value very close to direct stress measurements on substrate-adhering cells \cite{balaban2001force,tan2003cells} considering the $\sim 2\mu$m$^2$ cross-sectional size of epithelial focal adhesions \cite{dolat2014septins,Kuo2014}.
 
Figure~\ref{fig:Cell Results}A shows that the apical, lateral, and basal cross sections of the resulting 3D shell shape capture the qualitative shapes observed in experiment with their distinctive differences. Moreover, Fig.~\ref{fig:Cell Results}B demonstrates excellent quantitative agreement of the shape profile $\epsilon_p(z)$  with the convolution microscopy data and confirms the boundary-layer-and-plateau character of the deformations. 
In particular, $\epsilon_p$ is, at any $z$,  significantly greater than the largest allowable 2D value $\epsilon_{p,c}$, and its $z$-averaged value (close to the numbers obtained from lower-resolution images) far exceeds $\epsilon_{p,c}$, thus explaining the discrepancy in Fig.~\ref{fig:2Dresults}C.
Note that, while the actin bundles locally cause the prominent basal boundary layer, the elevated plateau value would be impossible without the presence of the apical actin.

\begin{figure}[t]
    \centering
    \includegraphics[width=0.48\textwidth]{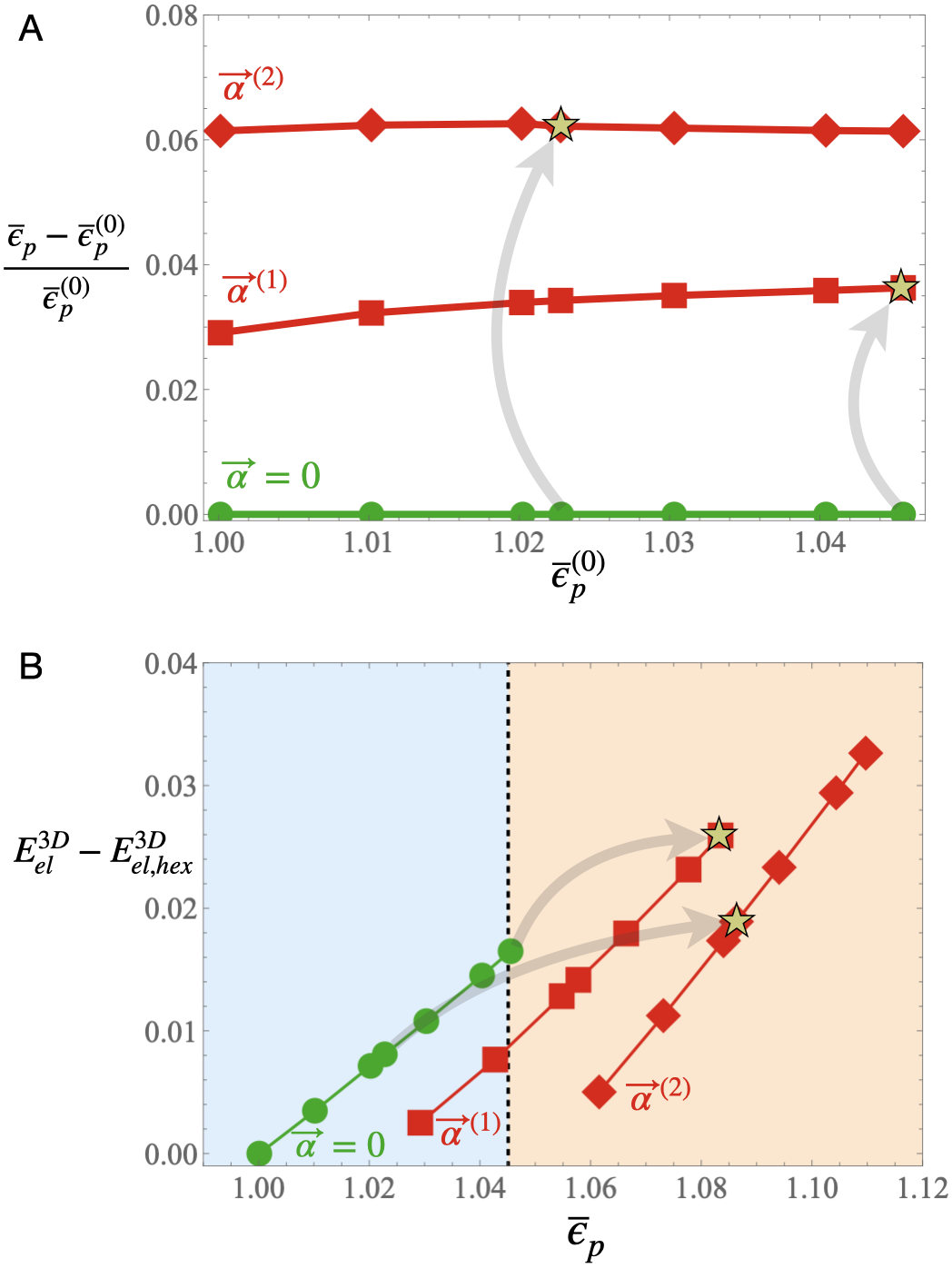}
    \caption{Deformation and mechanical energy. (A) Shell deformation caused by actin. Starting from the perimeter measures from passive confinement $\bar{\epsilon}_p^{(0)}$ compatible with 2D rigidity (green), two different strengths $\vec{\alpha}^{(1,2)} = (\alpha_A^{(1,2)}, \alpha_B^{(1,2)})$ of actin are applied, resulting in nearly constant relative increase of the perimeter measure (red). The experimentally observed $\bar{\epsilon}_p$ is reached from the largest $\bar{\epsilon}_p^{(0)}$ and the average $\bar{\epsilon}_p^{(0)}$, respectively (gray arrows, stars). (B) Increase in elastic energy as the passive confinement shape of the shells is varied. Green: without actin ($\vec{\alpha}=0$). Red: actin strengths $\vec{\alpha}^{(1,2)}$ as in A. Stars indicate shapes reproducing experimental data, see Fig.~\ref{fig:Cell Results}B; the slope of the curves in these points determines the critical adhesion strength $\gamma_c$.
    }
    \label{fig:Perimeter Prediction}
\end{figure}
\subsection*{Elastic energy, tissue rigidity, and tissue integrity}
The shell model quantifies all stress and strain components of the shell, and thus assesses the relative importance of deformation modes for the overall mechanical energy of the system. All elastic energy contributions display a similar boundary-layer variation as the displacement variables, dominated by the basal contributions of azimuthal stress ($N_\phi \varepsilon_\phi$), see Supporting Fig.~7 (SI) for the details. What are the implications for tissue rigidity?

The loss of rigidity found in 2D vertex models 
can be understood from the total mechanical energy functional $\tilde{E}_{tot}$ of a cell, consisting (in its simplest form) of the sum of an elastic contribution and a contribution of cell-cell adhesion, both depending on cell perimeters $P$ only. For consistent comparison with 3D energies, we interpret these terms as depending on the mantle area of strictly $z$-uniform cylinders of height $\Lambda a$, and non-dimensionalize
 energies by the scale $\tilde{E}_{ref}=2\pi\Lambda a^2K_A$. Here, $K_A$ is the effective area extension modulus of the shell material, which can be related to the 3D shell material properties by $K_A=Yt/(1-\nu^2)$ \cite{landau2020theory}. When the derivative of $E_{tot}=\tilde{E}_{tot}/\tilde{E}_{ref}$ with respect to average perimeter vanishes, the system loses internal stress and thus rigidity. Using $\epsilon_p=\bar{\epsilon}_p$ defined in \eqref{epsilonp} as the perimeter variable, the total energy functional is
\begin{equation}
E_{tot}^{2D} = E_{el}^{2D} + E_{adh}^{2D} = \frac{1}{2}\left(\kappa\bar{\epsilon}_p - 1\right)^2 - \gamma \kappa \bar{\epsilon}_p\,,
\label{2denergyeps}    
\end{equation}
where $\kappa = P_{hex}/P_0 \approx 1.05008\dots$ is the constant ratio of the perimeters of equal-area regular hexagon and circle, and $\gamma=\Gamma/K_A$ is the ratio of specific adhesion energy $\Gamma$ to the area modulus \cite{Kim15,Kim19}. The condition $\partial E_{tot}^{2D}/\partial \bar{\epsilon}_p = 0$ results in a critical strength of adhesion $\gamma_c\approx 0.12$ beyond which the system loses rigidity \cite{Kim15}. Somewhat counterintuitively, but in line with the earliest results in such modeling \cite{Bi_2015,lawson2021jamming}, this means that an increase in cell-cell adhesion ultimately compromises tissue stiffness and integrity (loss-of-rigidity transition). 

Our 3D elastic shell formalism allows for quantitative comparison with this result: The elastic energy $E_{el}^{3D}$ is computed as outlined above, and the adhesion energy $E_{adh}^{3D}$ is the specific adhesion energy (still assumed uniform) multiplied by the mantle area of the 3D deformed cylinder, which likewise follows from the shell model results. The energies are parametrized by 
$\bar{\epsilon}_p$. In Fig.~\ref{fig:Perimeter Prediction}A, we quantify the effect of actin stress on the perimeter measure for two sets of actin strengths: $(\alpha_A^{(1)}, \alpha_B^{(1)})$ as obtained above and $(\alpha_A^{(2)}\approx 3.7\times 10^{-4}, \alpha_B^{(2)}\approx 6.9\times 10^{-4})$, following from repeating the fit to Fig.~\ref{fig:Perimeter Measurement.png}B, but setting $\bar{\epsilon}_p^{(0)} = (1+\epsilon_{p,c}
)/2$, making the more realistic assumption that the experimental samples on which Fig.~\ref{fig:Perimeter Measurement.png}B is based have average, rather than extreme, passive deformations. Consequently, the values $(\alpha_A^{(2)}, \alpha_B^{(2)})$ are larger, as more of the deformation is due to active stresses; however, the difference is not great indicating that the modeled active stresses are insensitive to the passive-stress conditions. Figure~\ref{fig:Perimeter Prediction}A shows that a given amount of actin stress increases $\bar{\epsilon}_p$ by a nearly constant percentage over the entire range of $\bar{\epsilon}_p^{(0)}$ allowed by rigidity in the 2D theory.

Unsurprisingly, these strongly deformed shapes have significantly larger elastic energy, as shown in  Fig.~\ref{fig:Perimeter Prediction}B, where the energy $E_{el,hex}^{3D}$ of an actin-less shell with the lowest $\bar{\epsilon}_p^{(0)}=1$ (regular hexagonal cross section) is used as a baseline. 
Note that the perimeter measures of nearly all of these shells are incompatible with the rigid range of the 2D theory (shaded blue in Figs.~\ref{fig:2Dresults}C and \ref{fig:Perimeter Prediction}B).
The elastic shell energy $E_{el}^{3D}$ explicitly obtained from 3D modeling increases with  
$\bar{\epsilon}_p$, and more strongly so for larger actin strength. This slope, combined with the $\bar{\epsilon}_p$ derivative of $E_{adh}^{3D}$, yields a critical cell-cell adhesion strength to undergo loss of rigidity. For the experimentally observed shapes in rigid 3D tissues (stars in Fig.~\ref{fig:Perimeter Prediction}), these values are $\gamma_c^{(1)}\approx 0.833$ and $\gamma_c^{(2)}\approx 0.848$, respectively, consistent with each other and dramatically larger than the 2D result (note that the definition of the quantity $\gamma$ is the same in 2D and 3D).

In addition to predicting realistic cell shapes in confluent epithelia, the shell theory therefore reveals a quite different view of the loss-of-rigidity transition: 
Tissues with realistic 3D elastic deformations can afford large cell-cell adhesion before losing rigidity, possibly more adhesion than can reasonably be present: Note that values on the order of $\gamma_c\sim 1$ imply that the adhesion energy balances energies on the order of $\tilde{E}_{ref}$, which is the energy scale of a shell with ${\cal O}(1)$ strain throughout. Thus, the mechanics of 3D deformation protects the tissue against loss of stiffness, and allows for a better compromise of high tissue integrity (toughness) by adhesion and simultaneously high rigidity (stiffness).

\section*{Conclusions}
The benefits of the 3D continuum shell approach to cell and tissue mechanics span across length scales and across applications: The model takes into account biochemical and morphological data on scales below the single-cell size, resolving features of polarisation along the apical-basal axis such as mechanical boundary layers as well as their dependence on parameters including actomyosin contractility. The connection between this information and the model predictions is immediate and quantitative, the former providing direct boundary conditions for the latter. In acknowledging the active character of the material, the approach also allows for new insights into the overall mechanical state of the tissue, its robustness, stiffness, and toughness. In particular, we conclude that 2D vertex models strongly underestimate the resilience of tissues against floppy modes of deformation, suggesting that even very strong cell-cell adhesion is compatible with rigid cell mechanics and finite tissue shear moduli. An extension of the present mean-field single-shell model to an explicit modeling of neighboring cells is straightforward (through matching conditions), will allow further reduction of free parameters, and will provide a fresh look at the role of disorder in the cellular pattern of confluent epithelial tissues. 

\section*{Materials and Methods}
\subsection*{Antibodies and reagents}
RR1 antibody to dog E-cadherin \cite{gumbiner1986functional} was provided by Barry Gumbiner (University of Virginia).  Antibodies to $\beta$-catenin (catalog no.\ 7963) were purchased from Santa Cruz Biotechnology.  Anti-ZO-1 monoclonal R40.76 originally described by Anderson et al.\ \cite{anderson1988characterization}  was obtained from the Developmental Studies Hybridoma Bank, created by the NICHD of the NIH and maintained at The University of Iowa, Department of Biology, Iowa City, IA 52242. Rhodamine phalloidin (catalog no.\ 4095645) was purchased from Bachem. Cytochalasin D (catalog no.\ 1233) and Blebbistatin (catalog no.\ 1852) were purchased from Tocris. 

\subsection*{Cell line}
 MDCK II cells \cite{hansson1986two} were maintained in MEM/Earle’s balanced salt solution (EBSS) supplemented with 25 mM Hepes and 10.
 
\subsection*{Immunofluorescence}  
MDCK II cells grown on Transwell-Clear filters (Corning) for 10 d were rinsed twice in EBSS and fixed in $1\%$ paraformaldehyde in 150 mM NaCl, 20 mM Hepes, pH 7.8, at 4°C for 90 min. The reaction was quenched and the cells permeabilized with 50 mM Tris in staining buffer 
($0.1\%$ Triton X-100, 100 mM NaCl, 20 mM Hepes, pH 7.8) for 1 h. After rinsing in staining buffer, the cells were incubated with primary antibodies in staining buffer overnight. After rinsing in staining buffer three times, the cells were incubated in secondary antibodies for 90 min. The cells were rinsed again three times and poststain fixed with $0.2\%$ paraformaldehyde in staining buffer. Finally, the cells were incubated with fluorescently labeled phalloidin for 60 min and then rinsed twice with staining buffer. Transwell filters were excised and mounted on glass slides using ProLong Gold antifade (Invitrogen).
 
\subsection*{Imaging}
Low magnification wide-field images were collected using a Plan Apochromat 20×/0.8 objective (Zeiss) on an AxioImager.Z2m microscope equipped with Apotome.2 (Carl Zeiss), an X-cite 120 LED light source (Lumen Dynamics), and a 4K ORCA-Flash4.0 V2 digital CMOS camera with $6.5\mu$m$\times 6.5\mu$m pixel size (Hamamatsu Photonics) all driven by Zen2 Zeiss Software.

Optical z slices in 200-nm steps were collected with a microscope (1X71; Olympus) attached to a 1K × 1K charge-coupled device camera using a 60× objective (NA 1.42) with a 1.6× auxiliary magnification. All images were deconvolved using DeltaVision software (Applied Precision). Z stack projections were generated from deconvolved slices using the maximum intensity criteria. Composite images were generated using FIJI \cite{schindelin2012fiji}. 
 
\subsection*{Image Analysis}
The fluorescent signal of $\alpha$-actinin and $\beta$-catenin is the primary data to extract the morphology of the cells. In low-magnification (20x) images, the depth resolution is limited and we aim for extracting the 2D polygonal outlines of the cells. Results for $\bar{\epsilon}_p$ from both fluorescence channels are consistent with each other; we focus on $\alpha$-actinin here. We used Cellpose to segment the $N\sim 2,000$ cells in each image  \cite{pachitariu2022cellpose,stringer2021cellpose}; the MDCK cell histogram of Fig.~\ref{fig:2Dresults}C contains information from $\sim 12,000$ cells. A custom MATLAB algorithm was written to automatically (i) skeletonize the resulting boundaries, and (ii) quantify geometrical and topological data (cell perimeters, areas, and neighbor numbers). We take into account that perimeter outlines of cells are not made of straight lines, using a cubic spline fit to extract true lengths. The cell boundaries are interpreted as ideally one-dimensional, so that the areas $A_i$ add up to the total sample area. Thus, we compute $\epsilon_p$ for each sample. The coefficient of variation (polydispersity of areas) is $c_A\approx 0.311 \pm 0.002 $  consistently over all samples, within the range of accurate application of the 2D theory determining $\epsilon_{p,c}$ \cite{Kim19}.

\subsection*{Equations of elastic shell deformation}
We use the following set of equations describing 3D deformations of a thin elastic shell of cross-sectional area $\pi a^2$ (cf.\ \cite{flügge2013stresses,timoshenko1959theory,leissa1973vibration}). Here, $p_r$ and $p_\phi$ are the radial and azimuthal stresses on the shell caused by neighboring cells (stresses are non-dimensionalized by $Y$). $\nu$ is Poission's ratio, and derivatives with respect to 
$z/a$ and $\phi$ are designated by $\prime$ and $\dot{\ }$\,, respectively. 

\begin{align}
 &\frac{(1-\nu)}{2} \ddot{u}+\nu w'+\frac{(1+\nu)}{2}  \dot{v}'+u'' + \nonumber\\ &\frac{1}{12} t^2 \left( \frac{(1-\nu)}{2}  \ddot{u} + \frac{(1 - \nu)}{2}  \ddot{w}' - w''' \right)  =  0  \label{shellequations1} \\ 
& \dot{w} + \ddot{v} + \frac{(1 + \nu)}{2}  \dot{u}' + \frac{(1 - \nu)}{2}  v'' + \nonumber\\ &\frac{1}{12} t^2 \left( \frac{3(1 - \nu)}{2}  v'' - \frac{(3 - \nu)}{2}  \dot{w}'' \right) + \frac{(1 - \nu^2)}{t} p_\phi  =   0 \label{shellequations2} \\ 
& w + \dot{v} + \nu u' + \nonumber\\ 
&\frac{1}{12} t^2 \left( w + 2 \ddot{w} + \ddot{\ddot{w}} + \frac{(1 - \nu)}{2}  \ddot{u}' - \frac{(3 - \nu)}{2}  \dot{v}'' + 2 \ddot{w}'' - u''' + w'''' \right) \nonumber\\
& - \frac{(1 - \nu^2)}{t} p_r  =   0
\label{shellequations3}
\end{align}
A Fourier decomposition of the variables, e.g.\ $w (z,\phi) = \sum w_n(z) \cos(2n\phi)$, transforms this system to ODEs. Eliminating $u_n,v_n$ results in an $8^{th}$-order equation for $w_n(z)$, whose solution is of the form
\begin{equation}
w_n(z) = w_{n,p}(z) + \sum_{k=1}^{8} c_{nk}e^{\lambda_{nk}z}\,,
\label{wnofz}    
\end{equation}
where the particular solution $w_{n,p}$ is determined by the neighbor constraint stresses $p_r, p_\phi$. Note that, even though the neighbor stresses are unbiased along $z$, $w_{n,p}$ is not exactly uniform in $z$ (cf.\ Fig.~\ref{fig:Cell Results}B), as the displacement boundary conditions at $z=0$ and $z=\Lambda a$ are not symmetric. The homogeneous part of the solution consists of exponentials with complex eigenvalues $\lambda_{nk}$ determined from the characteristic equation of the ODE, and complex coefficients $c_{nk}$ determined from the actin stress boundary conditions.

\subsection*{Boundary conditions}
\eqref{shellequations1}- \eqref{shellequations3} are subject to eight boundary conditions. Basally ($z=0$) these are $u=v=0$ (no substrate deformation), $M_z=0$ (no net bending of the cylinder), and $S_z=S_z^B$ for the active radial shear stress resultants from actin bundles as detailed in the main text. At the free apical end, ($z=\Lambda a$), we have $N_z=0$ (no normal stress), $M_z=0$, and the radial and azimuthal shear resultants determined by the isotropic actin cortex strength $S_z=S_z^A$, $T_z=T_z^A$. These components are projective components of the uniform stress resultant locally normal to the polygonal cylinder outline (see SI).

\subsection*{3D Elastic Energies}
The (dimensional) elastic shell energy $\tilde{E}^{3D}_{el}$ is the sum of contributions from all strain modes, but is strongly dominated by the azimuthal term composed of the azimuthal strain 
\begin{equation}
\varepsilon_\phi = \dot{v} + w
    \label{epsphi}
\end{equation}
and the azimuthal stress resultant
\begin{equation}
N_\phi= \frac{Y t}{1-\nu^2} \left[\dot{v} + w + \nu u^\prime + \frac{t^2}{12}\left(w+\ddot{w}\right) \right]\,,
\label{nphi}
\end{equation}
leading to the explicit computation of the dimensionless 3D mechanical energy as an integral over the shell area,
\begin{equation}
E^{3D}_{el} = \frac{1}{\tilde{E}_{ref} }\int\!\!\!\int N_\phi \varepsilon_\phi \,ds \,dz\,,
    \label{eel}
\end{equation}
with $ds$ as the infinitesimal perimeter variable. This energy is plotted in Fig.~\ref{fig:Perimeter Prediction}B vs.\
the average perimeter measure $\bar{\epsilon}_p$, which is likewise obtained explicitly from the 3D shape of deformation solutions. At a fixed $z$, the perimeter is
\begin{equation}
P(z) =\int ds = a \int \left( \left(1+w\right)^2+\dot{w}^2\right)^{1/2} d\phi\,
    \label{perimeterfromw}
\end{equation}
In exact analogy to \eqref{epsilonp}, $\epsilon_p(z) = P(z)/P_{hex}$, and $z$-averaging obtains $\bar{\epsilon}_p$.

\section*{Disclaimer}
This work was prepared while Vivian Tang was employed at the University of Illinois, Urbana-Champaign. The opinions expressed in this article are the author's own and do not reflect the view of the National Institutes of Health, the Department of Health and Human Services, or the United States government. 

\section*{Acknowledgments}
The authors are grateful for stimulating discussions with Richard Carthew, Sangwoo Kim, and Vinothan Manoharan. We thank Ran Yang and Eva Sue at University of Illinois at Urbana-Champaign for help acquiring data. M.Z.N. and J.M.R. thank the Northwestern University Quantitative Biology Center for initial funding under grant $\#$60050530. Support from the National Institutes of Health under R01-DK098398 to V.W.T. and R01-GM106106 to W.M.B. is gratefully acknowledged.

\bibliographystyle{unsrt}  
\bibliography{Reference}  

\begin{thebibliography}{10}

\bibitem{Staple10}
D.~B. Staple, R.~Farhadifar, J.-C. R{\"o}per, B.~Aigouy, S.~Eaton, and F.~J{\"u}licher.
\newblock Mechanics and remodelling of cell packings in epithelia.
\newblock {\em The European Physical Journal E}, 33(2):117--127, Oct 2010.

\bibitem{Bi_2015}
Dapeng Bi, J.~H. Lopez, J.~M. Schwarz, and M.~Lisa Manning.
\newblock A density-independent rigidity transition in biological tissues.
\newblock {\em Nature Physics}, 11(12):1074–1079, September 2015.

\bibitem{Kim15}
Sangwoo Kim and Sascha Hilgenfeldt.
\newblock Cell shapes and patterns as quantitative indicators of tissue stress in the plant epidermis.
\newblock {\em Soft Matter}, 11:7270--7275, 2015.

\bibitem{kim2021embryonic}
Sangwoo Kim, Marie Pochitaloff, Georgina~A Stooke-Vaughan, and Otger Camp{\`a}s.
\newblock Embryonic tissues as active foams.
\newblock {\em Nature physics}, 17(7):859--866, 2021.

\bibitem{banavar2021mechanical}
Samhita~P Banavar, Emmet~K Carn, Payam Rowghanian, Georgina Stooke-Vaughan, Sangwoo Kim, and Otger Camp{\`a}s.
\newblock Mechanical control of tissue shape and morphogenetic flows during vertebrate body axis elongation.
\newblock {\em Scientific reports}, 11(1):8591, 2021.

\bibitem{Kim19}
Sangwoo Kim and Sascha Hilgenfeldt.
\newblock A simple landscape of metastable state energies for two-dimensional cellular matter.
\newblock {\em Soft Matter}, 15:237--242, 2019.

\bibitem{wang2020anisotropy}
Xun Wang, Matthias Merkel, Leo~B Sutter, Gonca Erdemci-Tandogan, M~Lisa Manning, and Karen~E Kasza.
\newblock Anisotropy links cell shapes to tissue flow during convergent extension.
\newblock {\em Proceedings of the National Academy of Sciences}, 117(24):13541--13551, 2020.

\bibitem{thomas2023shape}
Evan~C Thomas and Sevan Hopyan.
\newblock Shape-driven confluent rigidity transition in curved biological tissues.
\newblock {\em Biophysical Journal}, 122(21):4264--4273, 2023.

\bibitem{Kim22}
Sangwoo Kim and Sascha Hilgenfeldt.
\newblock Structural measures as guides to ultrastable states in overjammed packings.
\newblock {\em Phys. Rev. Lett.}, 129:168001, Oct 2022.

\bibitem{anderson1988characterization}
James~Melvin Anderson, Bruce~R Stevenson, Lynne~A Jesaitis, Daniel~A Goodenough, and Mark~S Mooseker.
\newblock Characterization of zo-1, a protein component of the tight junction from mouse liver and madin-darby canine kidney cells.
\newblock {\em The Journal of cell biology}, 106(4):1141--1149, 1988.

\bibitem{schulze2017elastic}
KD~Schulze, SM~Zehnder, JM~Urue{\~n}a, T~Bhattacharjee, WG~Sawyer, and TE~Angelini.
\newblock Elastic modulus and hydraulic permeability of mdck monolayers.
\newblock {\em Journal of Biomechanics}, 53:210--213, 2017.

\bibitem{tang2013fsgs3}
Vivian~W Tang and William~M Brieher.
\newblock Fsgs3/cd2ap is a barbed-end capping protein that stabilizes actin and strengthens adherens junctions.
\newblock {\em Journal of Cell Biology}, 203(5):815--833, 2013.

\bibitem{lecuit2011force}
Thomas Lecuit, Pierre-Fran{\c{c}}ois Lenne, and Edwin Munro.
\newblock Force generation, transmission, and integration during cell and tissue morphogenesis.
\newblock {\em Annual review of cell and developmental biology}, 27(1):157--184, 2011.

\bibitem{heisenberg2013forces}
Carl-Philipp Heisenberg and Yohanns Bella{\"\i}che.
\newblock Forces in tissue morphogenesis and patterning.
\newblock {\em Cell}, 153(5):948--962, 2013.

\bibitem{tang2012alpha}
Vivian~W Tang and William~M Brieher.
\newblock $\alpha$-actinin-4/fsgs1 is required for arp2/3-dependent actin assembly at the adherens junction.
\newblock {\em Journal of Cell Biology}, 196(1):115--130, 2012.

\bibitem{saraswathibhatla2020tractions}
Aashrith Saraswathibhatla and Jacob Notbohm.
\newblock Tractions and stress fibers control cell shape and rearrangements in collective cell migration.
\newblock {\em Physical Review X}, 10(1):011016, 2020.

\bibitem{jetta2023epithelial}
Deekshitha Jetta, Tasnim Shireen, and Susan~Z Hua.
\newblock Epithelial cells sense local stiffness via piezo1 mediated cytoskeletal reorganization.
\newblock {\em Frontiers in Cell and Developmental Biology}, 11:1198109, 2023.

\bibitem{martin2009pulsed}
Adam~C Martin, Matthias Kaschube, and Eric~F Wieschaus.
\newblock Pulsed contractions of an actin--myosin network drive apical constriction.
\newblock {\em Nature}, 457(7228):495--499, 2009.

\bibitem{smutny2010myosin}
Michael Smutny, Hayley~L Cox, Joanne~M Leerberg, Eva~M Kovacs, Mary~Anne Conti, Charles Ferguson, Nicholas~A Hamilton, Robert~G Parton, Robert~S Adelstein, and Alpha~S Yap.
\newblock Myosin ii isoforms identify distinct functional modules that support integrity of the epithelial zonula adherens.
\newblock {\em Nature cell biology}, 12(7):696--702, 2010.

\bibitem{alt2017vertex}
Silvanus Alt, Poulami Ganguly, and Guillaume Salbreux.
\newblock Vertex models: from cell mechanics to tissue morphogenesis.
\newblock {\em Philosophical Transactions of the Royal Society B: Biological Sciences}, 372(1720):20150520, 2017.

\bibitem{rozman2020collective}
Jan Rozman, Matej Krajnc, and Primo{\v{z}} Ziherl.
\newblock Collective cell mechanics of epithelial shells with organoid-like morphologies.
\newblock {\em Nature communications}, 11(1):3805, 2020.

\bibitem{flügge2013stresses}
W.~Fl{\"u}gge.
\newblock {\em Stresses in Shells}.
\newblock Springer Berlin Heidelberg, 2013.

\bibitem{klingner2014isotropic}
Christoph Klingner, Anoop~V Cherian, Johannes Fels, Philipp~M Diesinger, Roland Aufschnaiter, Nicola Maghelli, Thomas Keil, Gisela Beck, Iva~M Toli{\'c}-N{\o}rrelykke, Mark Bathe, et~al.
\newblock Isotropic actomyosin dynamics promote organization of the apical cell cortex in epithelial cells.
\newblock {\em Journal of Cell Biology}, 207(1):107--121, 2014.

\bibitem{boggon2002c}
Titus~J Boggon, John Murray, Sophie Chappuis-Flament, Ellen Wong, Barry~M Gumbiner, and Lawrence Shapiro.
\newblock C-cadherin ectodomain structure and implications for cell adhesion mechanisms.
\newblock {\em Science}, 296(5571):1308--1313, 2002.

\bibitem{ishiyama2012three}
Noboru Ishiyama and Mitsuhiko Ikura.
\newblock {\em The Three-Dimensional Structure of the Cadherin--Catenin Complex}, pages 39--62.
\newblock Springer Netherlands, Dordrecht, 2012.

\bibitem{chugh2017actin}
Priyamvada Chugh, Andrew~G Clark, Matthew~B Smith, Davide~AD Cassani, Kai Dierkes, Anan Ragab, Philippe~P Roux, Guillaume Charras, Guillaume Salbreux, and Ewa~K Paluch.
\newblock Actin cortex architecture regulates cell surface tension.
\newblock {\em Nature cell biology}, 19(6):689--697, 2017.

\bibitem{mahadevan2007persistence}
L~Mahadevan, A~Vaziri, and Moumita Das.
\newblock Persistence of a pinch in a pipe.
\newblock {\em Europhysics Letters}, 77(4):40003, 2007.

\bibitem{deguchi2005evaluation}
S~Deguchi, T~Ohashi, and M~Sato.
\newblock Evaluation of tension in actin bundle of endothelial cells based on preexisting strain and tensile properties measurements.
\newblock {\em Molecular \& Cellular Biomechanics}, 2(3):125, 2005.

\bibitem{kassianidou2015biomechanical}
Elena Kassianidou and Sanjay Kumar.
\newblock A biomechanical perspective on stress fiber structure and function.
\newblock {\em Biochimica et Biophysica Acta (BBA)-Molecular Cell Research}, 1853(11):3065--3074, 2015.

\bibitem{katoh1998isolation}
Kazuo Katoh, Yumiko Kano, Michitaka Masuda, Hirofumi Onishi, and Keigi Fujiwara.
\newblock Isolation and contraction of the stress fiber.
\newblock {\em Molecular Biology of the Cell}, 9(7):1919--1938, 1998.

\bibitem{balaban2001force}
Nathalie~Q Balaban, Ulrich~S Schwarz, Daniel Riveline, Polina Goichberg, Gila Tzur, Ilana Sabanay, Diana Mahalu, Sam Safran, Alexander Bershadsky, Lia Addadi, et~al.
\newblock Force and focal adhesion assembly: a close relationship studied using elastic micropatterned substrates.
\newblock {\em Nature cell biology}, 3(5):466--472, 2001.

\bibitem{tan2003cells}
John~L Tan, Joe Tien, Dana~M Pirone, Darren~S Gray, Kiran Bhadriraju, and Christopher~S Chen.
\newblock Cells lying on a bed of microneedles: an approach to isolate mechanical force.
\newblock {\em Proceedings of the National Academy of Sciences}, 100(4):1484--1489, 2003.

\bibitem{dolat2014septins}
Lee Dolat, John~L Hunyara, Jonathan~R Bowen, Eva~Pauline Karasmanis, Maha Elgawly, Vitold~E Galkin, and Elias~T Spiliotis.
\newblock Septins promote stress fiber--mediated maturation of focal adhesions and renal epithelial motility.
\newblock {\em Journal of Cell Biology}, 207(2):225--235, 2014.

\bibitem{Kuo2014}
Chiung~Wen Kuo, Di{-}Yen Chueh, and Peilin Chen.
\newblock Investigation of size--dependent cell adhesion on nanostructured interfaces.
\newblock {\em Journal of Nanobiotechnology}, 12(1):54, 2014.

\bibitem{landau2020theory}
Lev~D Landau, Evgeni{\i}~M Lifshitz, RJ~Atkin, and N~Fox.
\newblock The theory of elasticity.
\newblock In {\em Physics of Continuous Media}, pages 167--178. CRC Press, 2020.

\bibitem{lawson2021jamming}
Elizabeth Lawson-Keister and M~Lisa Manning.
\newblock Jamming and arrest of cell motion in biological tissues.
\newblock {\em Current Opinion in Cell Biology}, 72:146--155, 2021.

\bibitem{gumbiner1986functional}
Barry Gumbiner and Kai Simons.
\newblock A functional assay for proteins involved in establishing an epithelial occluding barrier: identification of a uvomorulin-like polypeptide.
\newblock {\em The Journal of cell biology}, 102(2):457--468, 1986.

\bibitem{hansson1986two}
Gunnar~C Hansson, Kai Simons, and Gerrit van Meer.
\newblock Two strains of the madin-darby canine kidney (mdck) cell line have distinct glycosphingolipid compositions.
\newblock {\em The EMBO journal}, 5(3):483--489, 1986.

\bibitem{schindelin2012fiji}
Johannes Schindelin, Ignacio Arganda-Carreras, Erwin Frise, Verena Kaynig, Mark Longair, Tobias Pietzsch, Stephan Preibisch, Curtis Rueden, Stephan Saalfeld, Benjamin Schmid, et~al.
\newblock Fiji: an open-source platform for biological-image analysis.
\newblock {\em Nature methods}, 9(7):676--682, 2012.

\bibitem{pachitariu2022cellpose}
Marius Pachitariu and Carsen Stringer.
\newblock Cellpose 2.0: how to train your own model.
\newblock {\em Nature methods}, 19(12):1634--1641, 2022.

\bibitem{stringer2021cellpose}
C.~Stringer, T.~Wang, M.~Michaelos, and M.~Pachitariu.
\newblock Cellpose: a generalist algorithm for cellular segmentation.
\newblock {\em Nature Methods}, 18(1):100--106, 2021.

\bibitem{timoshenko1959theory}
S.~Timoshenko and S.~Woinowsky-Krieger.
\newblock {\em Theory of Plates and Shells}.
\newblock Engineering mechanics series. McGraw-Hill, 1959.

\bibitem{leissa1973vibration}
A.W. Leissa.
\newblock {\em Vibration of Shells}.
\newblock NASA SP. Scientific and Technical Information Office, National Aeronautics and Space Administration, 1973.

\end{thebibliography}


\begin{thebibliography}{4}%
\makeatletter
\providecommand \@ifxundefined [1]{%
 \@ifx{#1\undefined}
}%
\providecommand \@ifnum [1]{%
 \ifnum #1\expandafter \@firstoftwo
 \else \expandafter \@secondoftwo
 \fi
}%
\providecommand \@ifx [1]{%
 \ifx #1\expandafter \@firstoftwo
 \else \expandafter \@secondoftwo
 \fi
}%
\providecommand \natexlab [1]{#1}%
\providecommand \enquote  [1]{``#1''}%
\providecommand \bibnamefont  [1]{#1}%
\providecommand \bibfnamefont [1]{#1}%
\providecommand \citenamefont [1]{#1}%
\providecommand \href@noop [0]{\@secondoftwo}%
\providecommand \href [0]{\begingroup \@sanitize@url \@href}%
\providecommand \@href[1]{\@@startlink{#1}\@@href}%
\providecommand \@@href[1]{\endgroup#1\@@endlink}%
\providecommand \@sanitize@url [0]{\catcode `\\12\catcode `\$12\catcode `\&12\catcode `\#12\catcode `\^12\catcode `\_12\catcode `\%12\relax}%
\providecommand \@@startlink[1]{}%
\providecommand \@@endlink[0]{}%
\providecommand \url  [0]{\begingroup\@sanitize@url \@url }%
\providecommand \@url [1]{\endgroup\@href {#1}{\urlprefix }}%
\providecommand \urlprefix  [0]{URL }%
\providecommand \Eprint [0]{\href }%
\providecommand \doibase [0]{https://doi.org/}%
\providecommand \selectlanguage [0]{\@gobble}%
\providecommand \bibinfo  [0]{\@secondoftwo}%
\providecommand \bibfield  [0]{\@secondoftwo}%
\providecommand \translation [1]{[#1]}%
\providecommand \BibitemOpen [0]{}%
\providecommand \bibitemStop [0]{}%
\providecommand \bibitemNoStop [0]{.\EOS\space}%
\providecommand \EOS [0]{\spacefactor3000\relax}%
\providecommand \BibitemShut  [1]{\csname bibitem#1\endcsname}%
\let\auto@bib@innerbib\@empty
\bibitem [{\citenamefont {Boudaoud}\ \emph {et~al.}(2014)\citenamefont {Boudaoud}, \citenamefont {Burian}, \citenamefont {Borowska-Wykręt},\ and\ \citenamefont {et~al.}}]{boudaoud2014fibriltool}%
  \BibitemOpen
  \bibfield  {author} {\bibinfo {author} {\bibfnamefont {A.}~\bibnamefont {Boudaoud}}, \bibinfo {author} {\bibfnamefont {A.}~\bibnamefont {Burian}}, \bibinfo {author} {\bibfnamefont {D.}~\bibnamefont {Borowska-Wykręt}},\ and\ \bibinfo {author} {\bibnamefont {et~al.}},\ }\bibfield  {title} {\bibinfo {title} {Fibriltool, an imagej plug-in to quantify fibrillar structures in raw microscopy images},\ }\href {https://doi.org/10.1038/nprot.2014.024} {\bibfield  {journal} {\bibinfo  {journal} {Nat Protoc}\ }\textbf {\bibinfo {volume} {9}},\ \bibinfo {pages} {457} (\bibinfo {year} {2014})}\BibitemShut {NoStop}%
\bibitem [{\citenamefont {Louveaux}\ and\ \citenamefont {Boudaoud}(2018)}]{louveaux2018fibriltool}%
  \BibitemOpen
  \bibfield  {author} {\bibinfo {author} {\bibfnamefont {M.}~\bibnamefont {Louveaux}}\ and\ \bibinfo {author} {\bibfnamefont {A.}~\bibnamefont {Boudaoud}},\ }\href {https://doi.org/10.5281/zenodo.2528872} {\bibinfo {title} {Fibriltool batch: an automated version of the imagej/fiji plugin fibriltool (version v1.0)}} (\bibinfo {year} {2018})\BibitemShut {NoStop}%
\bibitem [{\citenamefont {Wang}\ \emph {et~al.}(2020)\citenamefont {Wang}, \citenamefont {Merkel}, \citenamefont {Sutter}, \citenamefont {Erdemci-Tandogan}, \citenamefont {Manning},\ and\ \citenamefont {Kasza}}]{wang2020anisotropy}%
  \BibitemOpen
  \bibfield  {author} {\bibinfo {author} {\bibfnamefont {X.}~\bibnamefont {Wang}}, \bibinfo {author} {\bibfnamefont {M.}~\bibnamefont {Merkel}}, \bibinfo {author} {\bibfnamefont {L.~B.}\ \bibnamefont {Sutter}}, \bibinfo {author} {\bibfnamefont {G.}~\bibnamefont {Erdemci-Tandogan}}, \bibinfo {author} {\bibfnamefont {M.~L.}\ \bibnamefont {Manning}},\ and\ \bibinfo {author} {\bibfnamefont {K.~E.}\ \bibnamefont {Kasza}},\ }\bibfield  {title} {\bibinfo {title} {Anisotropy links cell shapes to tissue flow during convergent extension},\ }\href@noop {} {\bibfield  {journal} {\bibinfo  {journal} {Proceedings of the National Academy of Sciences}\ }\textbf {\bibinfo {volume} {117}},\ \bibinfo {pages} {13541} (\bibinfo {year} {2020})}\BibitemShut {NoStop}%
\bibitem [{\citenamefont {Fl{\"u}gge}(2013)}]{flügge2013stresses}%
  \BibitemOpen
  \bibfield  {author} {\bibinfo {author} {\bibfnamefont {W.}~\bibnamefont {Fl{\"u}gge}},\ }\href {https://books.google.com/books?id=tFXuCAAAQBAJ} {\emph {\bibinfo {title} {Stresses in Shells}}}\ (\bibinfo  {publisher} {Springer Berlin Heidelberg},\ \bibinfo {year} {2013})\BibitemShut {NoStop}%
\end{thebibliography}%

\end{document}


\title{Supporting Information to \\ Cell Deformation Signatures along the Apical-Basal Axis: A 3D Continuum Mechanics Shell Model}

\author{Jairo M. Rojas}
\affiliation{Department of Physics, The Grainger College of Engineering, University of Illinois, Urbana-Champaign, Illinois 61801, USA }
\affiliation{Mechanical Science and Engineering, The Grainger College of Engineering, University of Illinois, Urbana-Champaign, Illinois 61801, USA}
\author{Mayisha Z. Nakib}
\affiliation{Department of Physics, University of Illinois, Urbana-Champaign, Illinois 61801, USA }
\affiliation{Mechanical Science and Engineering, University of Illinois, Urbana-Champaign, Illinois 61801, USA}
\author{Vivian W. Tang}
\author{William M. Brieher}
\affiliation{Department of Cell and Developmental Biology, University of Illinois, Urbana-Champaign, Illinois 61801, US}
\author{Sascha Hilgenfeldt}
\affiliation{Department of Physics, University of Illinois, Urbana-Champaign, Illinois 61801, USA }
\affiliation{Mechanical Science and Engineering, University of Illinois, Urbana-Champaign, Illinois 61801, USA}

\maketitle

\subsection{Deformations occur consistently across epithelial cell samples}
\begin{figure}[H]
  \centering
  \includegraphics[width=12cm]{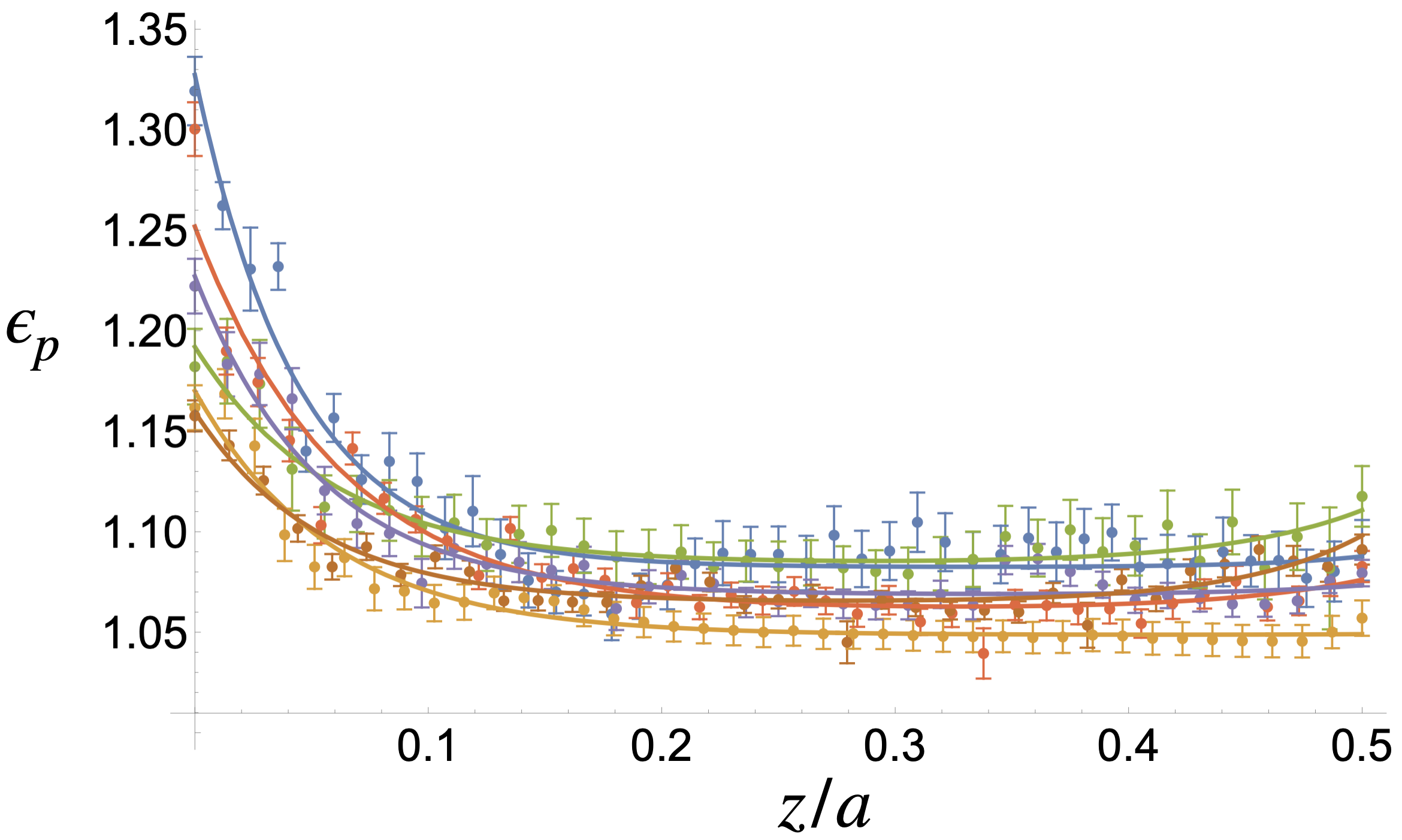} 
  \caption{Measured $z$-dependence of the perimeter measure $\epsilon_p$ for six individual MDCK cell samples (colors), each normalized by height to match the average height $\Lambda a$ (actual height variation between samples is only $\pm 0.5\mu$m). Each data point is the average of all cells in the 100x magnification sample whose perimeter could be quantified in all slices of the deconvolution microscopy stack; error bars are standard deviations. Lines are fits to the individual sample data with exponential boundary layers basally (dominant) and apically (weak).}
  \label{epszacrosssamples}
\end{figure}

The main text focuses on presenting quantitative results for the 3D shape of cells averaged over a series of samples taken at 100x magnification. While we cannot achieve the $z$-resolution necessary for such a detailed analysis in the much larger samples taken at 20x magnification (1000-2000 cells), we here report on the consistency of the results between individual cell samples imaged at the higher magnification (15-20 cells each, under identical experimental conditions sample to sample). Figure~\ref{epszacrosssamples} shows that the characteristic shape of cells as quantified by $\epsilon_p(z)$ is preserved across samples, and the standard deviations indicate very similar cell-to-cell variations. While different samples tend to have slightly different actomyosin contractility, and thus lengths of perimeter deformation that differ by up to $\sim 10\%$, the boundary-layer-and-plateau structure is preserved. Exponential fits to the boundary layers result in effective widths that vary little across samples (mean width: 
$\approx 0.055a$, standard deviation $\approx 0.006a$). Thus, the mean over all samples presented in the main text does not wash out or alter features seen in individual samples where cells are in direct mechanical contact with each other.

\subsection{Deformations occur consistently for individual cells}
\begin{figure}[H]
  \centering
  \includegraphics[width=12cm]{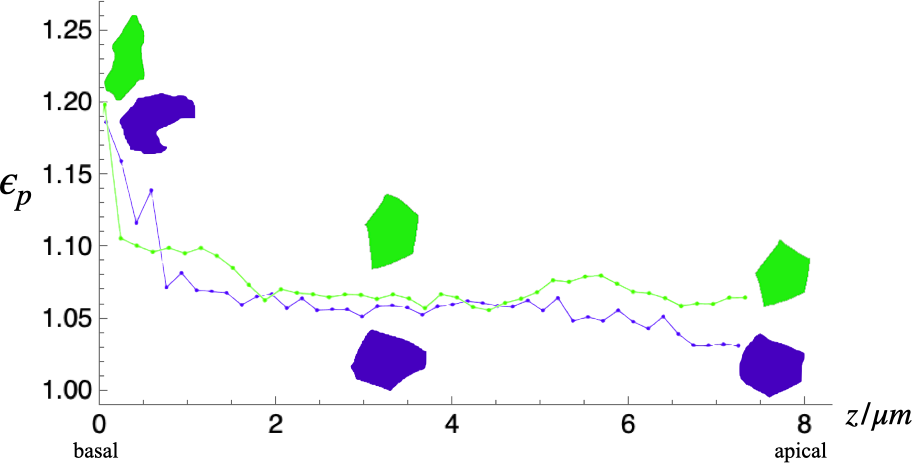} 
  \caption{Traces of $\epsilon_p(z)$ for two individual cells, showing perimeter measure evaluations from each individual microscopic slice. Despite marked differences in individual cross-sectional shape, the perimeter measures behave very similarly.}
  \label{epsindividual}
\end{figure}

Our work models a prototypical cell as an approach to quantifying deformation morphology, while neighboring cells are assumed to have an overall averaged effect on it (mean-field model). Therefore, our shell model exhibits symmetry with respect to the $\phi=\pi/2$ line as well as a hexagonal coordination number of the shell, whereas in a tissue sample, individual cells lack symmetries and can have different topologies (number of neighbors). Nevertheless, when we quantify perimeters of individual cells as a function of $z$ from the deconvolution microscopy stacks, the majority of their $\epsilon_p(z)$ shows a strong resemblance to the sample-averaged quantity (Fig.~\ref{epsindividual}), indicating that the shell deformation effects are experienced similarly by each individual cell. 

\subsection{Individual basal actin fiber bundle directions are strongly correlated with cell shape orientation}
\begin{figure}[H]
  \centering
  \includegraphics[width=12cm]{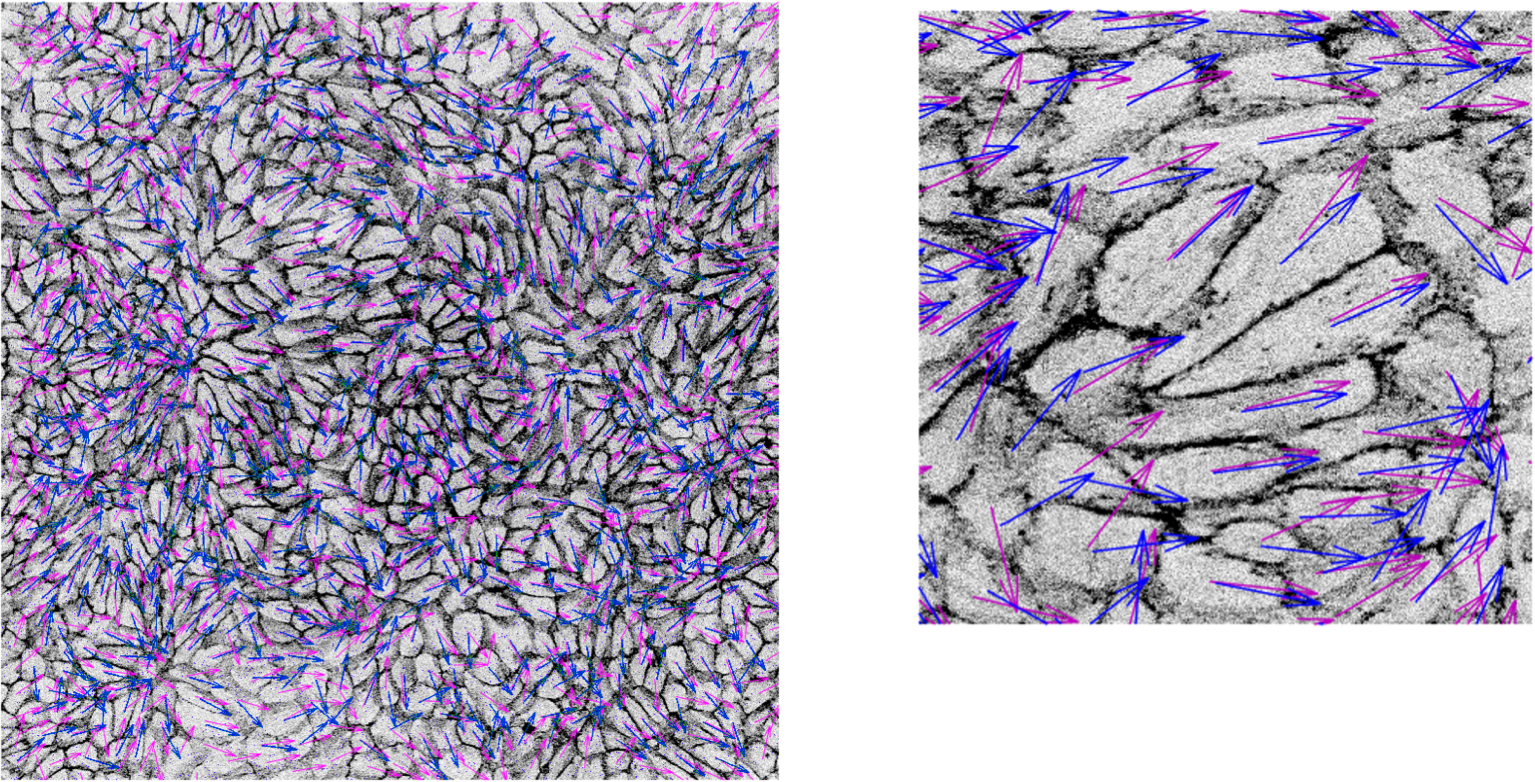} 
  \caption{Left: A sample of $\sim 2000$ MDCK cells, with arrows in magenta indicating fiber bundle orientation and arrows in purple indicating orientation of the cell cross section. Right: A close-up demonstrates that both orientation measures are closely aligned.}
  \label{cellvsbundleori}
\end{figure}

Our modeling crucially assumes that the main driver of 3D deformation is the presence of actin fibers and myosin motors, with contractility resulting from the action of the latter on the former. The actin present at the basal side plays a larger role than apical actin, both because of higher actin concentration basally and because of its organization into directional fiber bundles. If these assumptions are true, cell morphology should be at least partially determined by the direction of the actin bundles at the basal side. Using the FibrilTool ImageJ plug-in, it is straightforward to automatically identify actin bundle orientation in each cell \cite{boudaoud2014fibriltool}\cite{louveaux2018fibriltool}. Furthermore, segmented cells have a well-defined orientation, which can be defined by the angular direction of the major axis of their best-fit ellipse (here using corresponding functions of MATLAB's regionprops). As we do not need great $z$-resolution to resolve either fiber orientation or cell orientation, we obtain very good statistics from low-magnification (20x) samples, as shown in  Fig.~\ref{cellvsbundleori}. The figure shows that cell orientation indeed closely correlates with fiber bundle orientation: The  pull of the interior basal actin bundles of each cell  stretches it in the direction of the bundle orientation, supporting the elastic shell modeling results presented in the main text.

\subsection{Basal actin fiber bundle directions are uncorrelated within samples}
\begin{figure}[H]
  \centering
  \includegraphics[width=5cm]{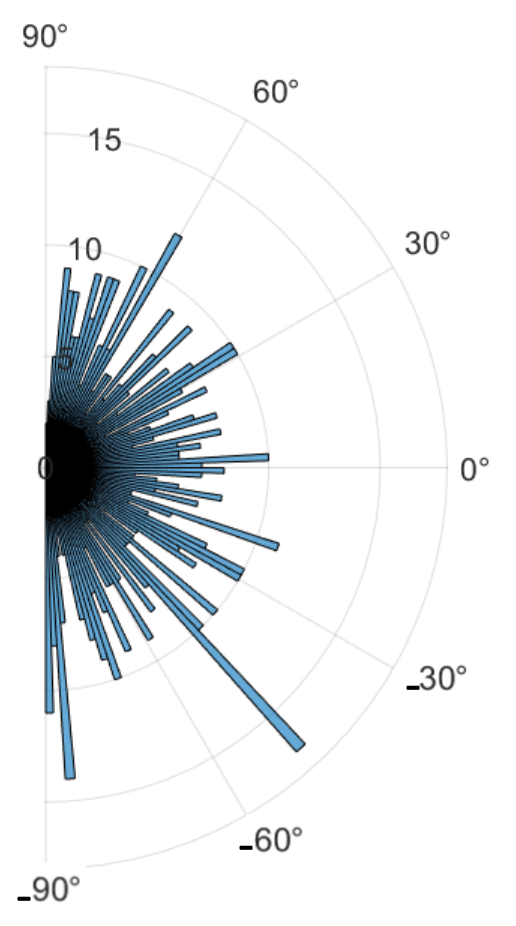} 
  \caption{Polar histogram quantifying the probability of fiber bundle orientation (relative to a chosen fixed direction on the substrate) by binned number counts. No systematic preferential orientation is evident.  }
  \label{bundleorientation}
\end{figure}

Even without 3D effects, perimeter values of (quasi-)2D domain structures can be much larger than the rigidity limit of 2D theory, if the domains are deformed systematically on the scale of the entire sample \cite{wang2020anisotropy}. A simple example of this is uniaxial stretch of a honeycomb pattern, which can increase the perimeters of the hexagon almost arbitrarily. However, such a deformation imparts large-scale anisotropy on the cells, i.e., the cells' major axes are then strongly aligned in the direction of the uniaxial stretch. Figure~\ref{bundleorientation} shows that the fiber bundle orientations (which, as was shown above, are a close surrogate for major axis orientation) are in fact uniformly distributed over all angles. Thus, we find no experimental evidence that large-scale anisotropy is behind the large $\bar{\epsilon}_p$ measures we observe.

\subsection{Drugs inhibiting actomyosin contractility decrease average cell perimeter measures}
\begin{figure}[H]
  \centering
  \includegraphics[width=7cm]{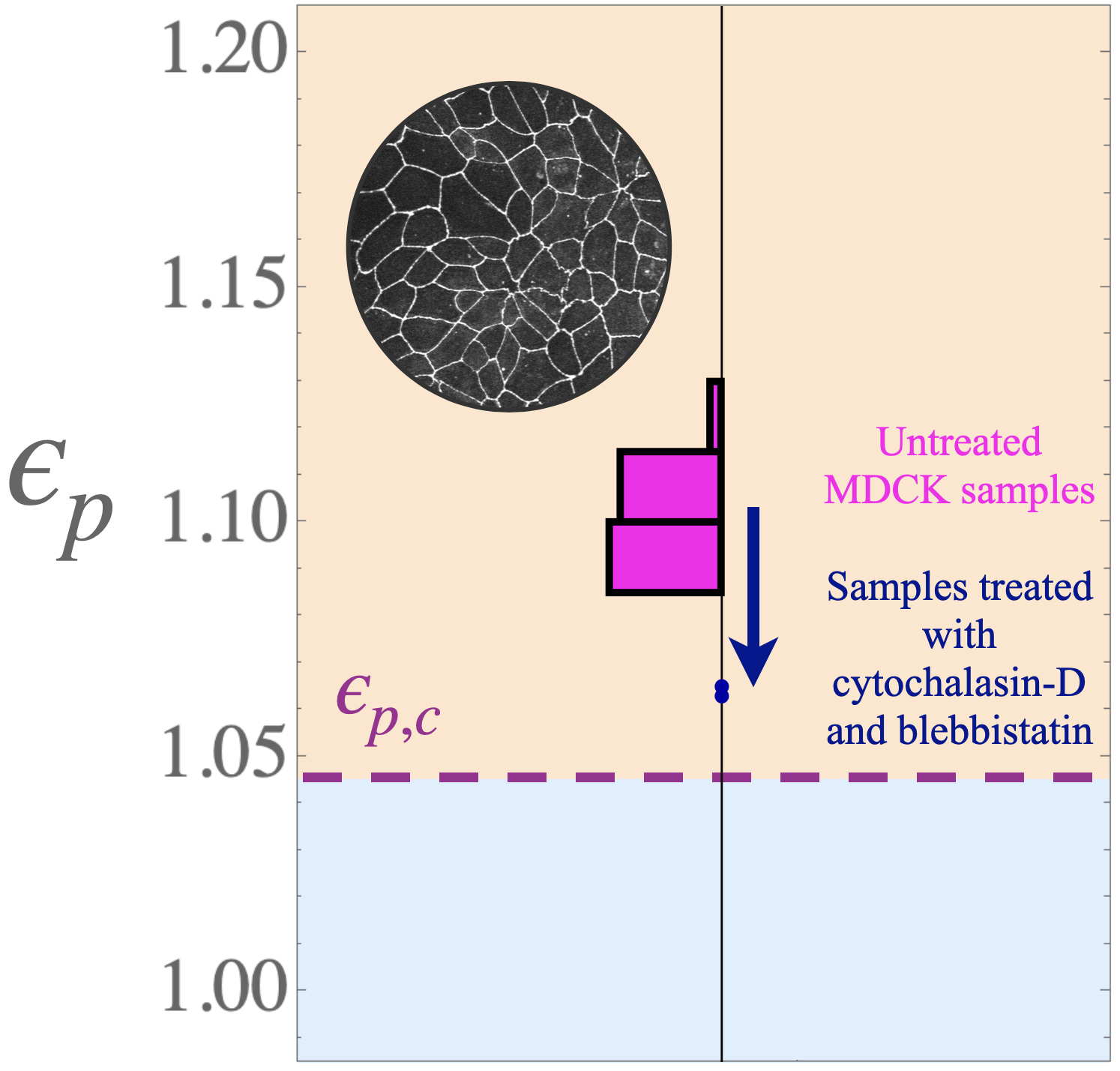} 
  \caption{Measured average perimeter measures $\bar{\epsilon}_p$ for twelve individual 20x MDCK samples (symbols) treated with cytochalasin-D and blebbistatin to disrupt actomyosin contractility. These values lie significantly below those of the untreated samples (histogram from Fig.1C of the main text).}
  \label{druggedsamples}
\end{figure}

Our modeling approach together with the experimental data we have obtained strongly suggests that the observed characteristic 3D deformations of cells are in large part a consequence of the active stresses present in the cell through actomyosin contractility. It is natural to ask whether inhibition of the cellular actomyosin machinery affects the cell shape changes. While a larger-scale quantitative investigation of such manipulation of the biological samples is projected, we here present preliminary results with MDCK samples treated with both $2.5\mu M$ cytochalasin-D (blocking actin polymerization) and $50\mu$M of the myosin inhibitor blebbistatin for 1 hour before fixing.
 Quantifying averaged perimeter measures $\bar{\epsilon}_p$ in such samples and superimposing them on the results for untreated epithelia shows a significant reduction of $\bar{\epsilon}_p$, though not low enough to be compatible with 2D theory (Fig.~\ref{druggedsamples}).

\subsection{Actin boundary conditions}
\begin{figure}[H]
  \centering
  \includegraphics[width=12cm]{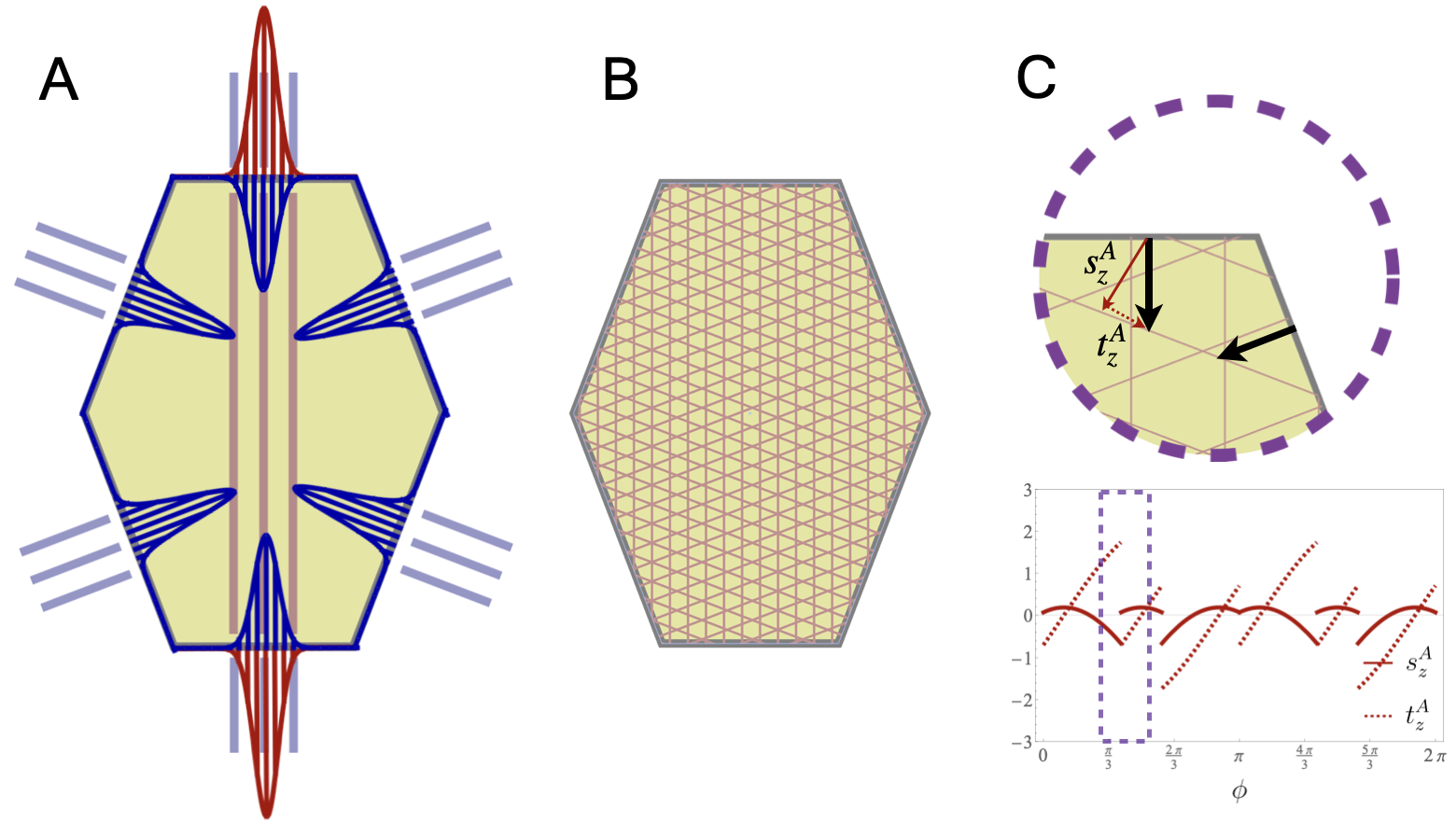} 
  \caption{(A) Sketch of basal polygonal cross section with radial shear stress resultants $S_z$ from internal (red) and external (blue) bundles represented by Gaussian functions of corresponding color. (B) Sketch of the apical polygonal cross section with a diffuse, isotropic actin cortex. (C) Close-up of one perimeter region of (B). The cortex exerts uniform normal force (black arrows), which is decomposed into its radial ($s_z^A$) and azimuthal ($t_z^A$) components (red arrows), whose magnitude depends on the angle between the orientation of the polygonal edge and the radial direction. The plot in (C) shows the resulting pattern of $s_z^A(\phi)$ and $t_z^A(\phi)$, with the dashed box corresponding to the portion of the perimeter in the close-up. See also Fig.3D of the main text.}
  \label{actinBCs}
\end{figure}

The boundary conditions exerted on the cylindrical shell basally ($z=0$) and apically ($z=\Lambda$) reflect the active stresses from actomyosin contractility. On the basal side, each cell has a strongly aligned internal actin bundle that traverses the cell radially, and consequently exerts radial stress resultants $s_z^{BI}$. We determine a typical width of these bundles in experiment and accordingly model the $\phi$-dependence of $s_z^{BI}$ as Gaussian peaks (at $\phi=\pi/2$ and $3\pi/2$, without loss of generality) of corresponding width $\Delta\phi$ (red functions sketched in Fig.~\ref{actinBCs}A). Neighboring cells have equivalent bundles of random orientation, so that in a mean-field model each face of the polygonal cylindrical shell is impacted by 1/3 of such a bundle (blue functions sketched in Fig.~\ref{actinBCs}A). Thus, we add $s_z^{BE}$ as six Gaussian peaks of 1/3 the height of the $s_z^{BI}$ peaks. The resulting $s_z^B=s_z^{BI}+s_z^{BE}$ has vanishing $n=0$ Fourier mode (the force pattern does not change cross-sectional area) and is finally normalized so that the root mean square of the $\phi$-dependent boundary-normal stress resultant is one (see Fig.3F of the main text). This normalization has to be performed with respect to the basal perimeter length $P_B= P(z=0)$ (cf.\ Eq.[10] of the main text), i.e., 
\[
\left[ \int (s_z^B)^2 ds\right]^{1/2}/P_B = 1\,.
\]

The prefactor $\alpha_B$ determining the actual strength of the basal stress resultant is found from a least-square fit to the observed deformations.

On the apical face of the cell, the actin cortex is diffuse and isotropic (Fig.~\ref{actinBCs}B). This implies uniform local normal forces and vanishing local tangential forces at every point of the apical circumference (see close-up in Fig.~\ref{actinBCs}C). However, the orientation of the edges of the polygonal circumference make non-zero angles with the $\phi$ direction, so that there are non-uniform radial shear stress resultants $S_z^A(\phi)$ as well as finite tangential shear stress resultants $T_z^A(\phi)$ \cite{flügge2013stresses} scaling with the cosine and sine of these angle orientation angles (Fig.~\ref{actinBCs}C). As the cross-sectional area on the apical side cannot change (without leading to a non-flat deformation of the entire tissue), the $n=0$ mode must be compensated by the action of the neighbors and is consequently subtracted. Normalizing the remaining stress resultants to a root mean square of one obtains $s_z^A(\phi)$ and $t_z^A(\phi)$, obeying
\[
\left[ \int (s_z^A)^2 + (t_z^A)^2  ds\right]^{1/2}/P_A = 1\,.
\]
The functional form of $s_z^A(\phi)$ and $t_z^A(\phi)$ gives rise to the pattern shown in Fig.3D of the main text, and the dimensionless strength $\alpha_A$ is again determined by the least-square fit to experimental data. 

\subsection{Energy contributions from different modes of elastic deformation}
\begin{figure}[H]
  \centering
  \includegraphics[width=12cm]{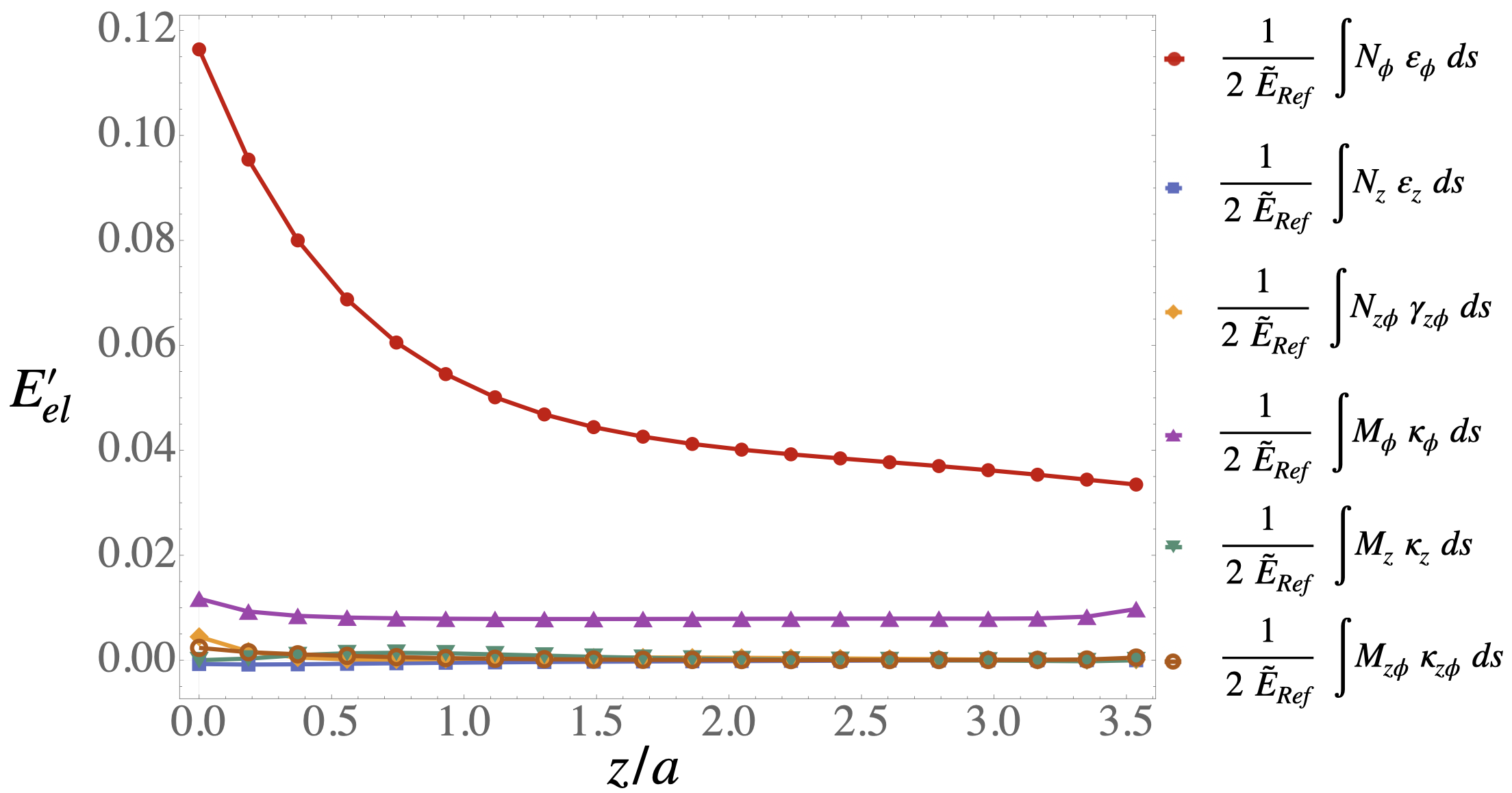} 
  \caption{Elastic energy components as a function of $z$, using the nomenclature of \cite{flügge2013stresses}: extensional strains $\varepsilon_\phi,\varepsilon_z$, shear strain $\gamma_{z\phi}$, curvature changes $\kappa_\phi, \kappa_z$, twist $\kappa_{z\phi}$, $N$ for stress resultants and $M$ for moment resultants, except denoting the axial coordinate by $z$ for consistency with the main text. The dominant azimuthal elastic energy (red) quantifies $E_{el}^{3D}$ in the main text.  }
  \label{elasticterms}
\end{figure}

From the elastic shell theory, all components of stress resultants and strains are quantitatively known. Elastic contributions to mechanical energy are obtained by multiplying corresponding components of stress resultants and strains, and integrating over the shell area. In Fig.~\ref{elasticterms}, we present these energy contributions per height along the $z$-axis (integration over the azimuthal line element $ds=a\left( \left(1+w\right)^2+\dot{w}^2\right)^{1/2} d\phi$ only). Energies are non-dimensional and normalized by $\tilde{E}_{Ref}$.  The elastic energy is strongly dominated by a single contribution, resulting from the product of the azimuthal stress resultant $N_\phi = \int \sigma_{\phi\phi} dr$ and the azimuthal normal strain $\varepsilon_\phi$. Furthermore, elastic energy from this term is concentrated in a boundary layer of width $\sim 0.1 a$ near the basal end. As all other contributions are much smaller, it is the azimuthal energy term $\int N_\phi \varepsilon_\phi  ds / (2 \tilde{E}_{ref})$ that we quantify as $E_{el}^{3D}$ in Fig.~5B of the main text.

\bibliography{Reference}